\documentclass{article}

\usepackage{microtype}
\usepackage{graphicx}
\usepackage{booktabs} 
\usepackage{setspace}

\usepackage{color}
\usepackage{enumerate}
\usepackage{amsmath}
\usepackage{amsthm}
\usepackage{amssymb}
\usepackage{url}
\usepackage{bbm}
\usepackage{subfig}
\usepackage{tikz}
\usetikzlibrary{shapes,arrows}
\usetikzlibrary{spy}
\usetikzlibrary{decorations.pathmorphing} 
\usetikzlibrary{fit}					
\usetikzlibrary{backgrounds}	

\usepackage{mathptmx} 
\usepackage{eurosym}
\usepackage{booktabs}
\usepackage{boxedminipage}
\usepackage{tabularx}
\usepackage{supertabular}
\usepackage[linesnumberedhidden,ruled,vlined]{algorithm2e}

\DeclareMathOperator*{\E}{\mathbb{E}}
\DeclareMathOperator*{\argmin}{arg\,min}
\DeclareMathOperator{\argmax}{arg\,max}

\newcommand{\timespan}[1]{\ensuremath{\delta(#1)}}
\newcommand{\soc}[2]{\ensuremath{SoC_{#1, #2}}}
\newcommand{\socmin}[1]{\ensuremath{\underline{SoC_{#1}}}}
\newcommand{\socmax}[1]{\ensuremath{\overline{SoC_{#1}}}}

\theoremstyle{remark}
\newtheorem{asu}{Assumption}

\usepackage{hyperref}
\usepackage{cleveref}

\usepackage[accepted]{icml2020}

\icmltitlerunning{A Deep Reinforcement Learning Framework for Continuous Intraday Market Bidding}

\begin{document}
	
	\twocolumn[
	\icmltitle{A Deep Reinforcement Learning Framework for Continuous Intraday Market Bidding}
	
	
	
	
	\begin{icmlauthorlist}
		\icmlauthor{Ioannis Boukas}{ulg}
		\icmlauthor{Damien Ernst}{ulg}
		\icmlauthor{Thibaut Th\'eate}{ulg}
		\icmlauthor{Adrien Bolland}{ulg}
		\icmlauthor{Alexandre Huynen}{engie}
		\icmlauthor{Martin Buchwald}{engie}
		\icmlauthor{Christelle Wynants}{engie}
		\icmlauthor{Bertrand Corn\'elusse}{ulg}
	\end{icmlauthorlist}
	
	\icmlaffiliation{ulg}{Department of Electrical Engineering and Computer Science,
		University of Li\`ege, Li\`ege, Belgium}
	\icmlaffiliation{engie}{Market Modeling and Market View, ENGIE, Brussels, Belgium}
	
	\icmlcorrespondingauthor{Ioannis Boukas}{ioannis.boukas@uliege.be}

	\icmlkeywords{Machine Learning, ICML}
	
	\vskip 0.5in
	]
	
	
	
	\printAffiliationsAndNotice{}  
	\begin{abstract}
		The large integration of variable energy resources is expected to shift a large part of the energy exchanges closer to real-time, where more accurate forecasts are available. In this context, the short-term electricity markets and in particular the intraday market are considered a suitable trading floor for these exchanges to occur. A key component for the successful renewable energy sources integration is the usage of energy storage. In this paper, we propose a novel modelling framework for the strategic participation of energy storage in the European continuous intraday market where exchanges occur through a centralized order book. The goal of the storage device operator is the maximization of the profits received over the entire trading horizon, while taking into account the operational constraints of the unit. The sequential decision-making problem of trading in the intraday market is modelled as a Markov Decision Process. An asynchronous distributed version of the fitted Q iteration algorithm is chosen for solving this problem due to its sample efficiency. The large and variable number of the existing orders in the order book motivates the use of high-level actions and an alternative state representation. Historical data are used for the generation of a large number of artificial trajectories in order to address exploration issues during the learning process. The resulting policy is back-tested and compared against a benchmark strategy that is the current industrial standard. Results indicate that the agent converges to a policy that achieves in average higher total revenues than the benchmark strategy. 
	\end{abstract}
	
	\section{Introduction}\label{sec: Introduction}
	The vast integration of renewable energy resources (RES) into (future) power systems, as directed by the recent worldwide energy policy drive \cite{EU2030}, has given rise to challenges related to the security, sustainability and affordability of the power system (``The Energy Trilemma''). The impact of high RES penetration on the modern short-term electricity markets has been the subject of extensive research over the last few years. Short-term electricity markets in Europe are organized as a sequence of trading opportunities where participants can trade energy in the day-ahead market and can later adjust their schedule in the intraday market until the physical delivery. Deviations from this schedule are then corrected by the transmission system operator (TSO) in real time and the responsible parties are penalized for their imbalances \cite{Meeus2017}.
	
	Imbalance penalties serve as an incentive for all market participants to accurately forecast their production and consumption and to trade based on these forecasts \cite{Scharff2016}. Due to the variability and the lack of predictability of RES, the output planned in the day-ahead market may differ significantly from the actual RES output in real time \cite{Karanfil2017}. Since the RES forecast error decreases substantially with a shorter prediction horizon, the intraday market allows RES operators to trade these deviations whenever an improved forecast is available \cite{Borggrefe2011}. As a consequence, intraday trading is expected to reduce the costs related to the reservation and activation of capacity for balancing purposes. The intraday market is therefore a key aspect towards the cost-efficient RES integration and enhanced system security of supply.
	
	Owing to the fact that commitment decisions are taken close to real time, the intraday market is a suitable market floor for the participation of flexible resources (i.e. units able to rapidly increase or decrease their generation/consumption). However, fast-ramping thermal units (e.g. gas power plants) incur a high cost when forced to modify their output, to operate in part load, or to frequently start up and shut down. The increased cost related to the cycling of these units will be reflected to the offers in the intraday market \cite{PerezArriaga2016}. Alternatively, flexible storage devices (e.g. pumped hydro storage units or batteries) with low cycling and zero fuel cost can offer their flexibility at a comparatively low price, close to the gate closure. Hence, they are expected to play a key role in the intraday market.
	
	\subsection{Intraday markets in Europe}
	
	In Europe, the intraday markets are organized in two distinct designs, namely auction-based or continuous trading.
	
	In auction-based intraday markets, participants can submit their offers to produce or consume energy at a certain time slot until gate closure. After the gate closure, the submitted offers are used to form the aggregate demand and supply curves. The intersection of the aggregate curves defines the clearing price and quantity \cite{Neuhoff2016}. The clearing rule is uniform pricing, according to which there is only one clearing price at which all transactions occur. Participants are incentivized to bid at their marginal cost since they are paid at the uniform price. This mechanism increases price transparency, although it leads to inefficiencies, since imbalances after the gate closure can no longer be traded \cite{Hagemann2013}.
	
	In continuous intraday (CID) markets, participants can submit at any point during the trading session orders to buy or to sell energy. The orders are treated according to the first come first served (FCFS) rule. A transaction occurs as soon as the price of a new ``Buy" (``Sell") order is equal or higher (lower) than the price of an existing ``Sell" (``Buy") order. Each transaction is settled following the pay-as-bid principle, stating that the transaction price is specified by the oldest order of the two present in the order book. Unmatched orders are stored in the order book and are accessible to all market participants. The energy delivery resolution offered by the CID market in Europe ranges between hourly, 30-minute and 15-minute products, and the gate closure takes place between five and 60 minutes before actual delivery. Continuous trading gives the opportunity to market participants to trade imbalances as soon as they appear \cite{Hagemann2013}. However, the FCFS rule is inherently associated with lower allocative inefficiency compared to auction rules. This implies that, depending on the time of arrival of the orders, some trades with a positive welfare contribution may not occur while others with negative welfare contribution may be realised \cite{Henriot2014}. It is observed that a combination of continuous and auction-based intraday markets can increase the market efficiency in terms of liquidity and market depth, and results in reduced price volatility \cite{Neuhoff2016}.
	
	In practice, the available contracts (``Sell" and ``Buy" orders) can be categorized into three types:
	\begin{itemize}
		\item The market order, where no price limit is specified (the order is matched at the best price)
		\item The limit order, which contains a price limit and can only be matched at that or at a better price
		\item The market sweep order, which is executed immediately (fully or partially) or gets cancelled.
	\end{itemize}
	Limit orders may appear with restrictions related to their execution and their validity. For instance, an order that carries the specification \textit{Fill or Kill} should either be fully and immediately executed or cancelled. An order that is specified as \textit{All or Nothing} remains in the order book until it is entirely executed \cite{Balardy2017}.
	
	The European Network Codes and specifically the capacity allocation and congestion management guidelines \cite{Meeus2017} (CACM GL) suggest that continuous trading should be the main intraday market mechanism. Complementary regional intraday auctions can also be put in place if they are approved by the regulatory authorities \cite{Meeus2017}. To that direction, the Cross-Border Intraday (XBID) Initiative \cite{XBID} has enabled continuous cross-border intraday trading across Europe. Participants of each country have access to orders placed from participants of any other country in the consortium through a centralized order book, provided that there is available cross-border capacity.
	
	\subsection{Bidding strategies in literature}
	
	The strategic participation of power producers in short-term electricity markets has been extensively studied in the literature. In order to co-optimise the decisions made in the sequential trading floors from day-ahead to real time the problem has been traditionally addressed using multi-stage stochastic optimisation. Each decision stage corresponds to a trading floor (i.e. day-ahead, capacity markets, real-time), where the final decisions take into account uncertainty using stochastic processes. In particular, the influence that the producer may have on the market price formation leads to the distinction between ``price-maker" and ``price-taker" and results in a different modelling of the uncertainty. 
	
	In \cite{Optimal1294977}, the optimisation of a portfolio of generating assets over three trading floors (i.e. the day-ahead, the adjustment and the reserves market) is proposed, where the producer is assumed to be a ``price-maker". The offering strategy of the producer is a result of the stochastic residual demand curve as well as the behaviour of the rest of the market players. On the contrary, a ``price-taker" producer is considered in \cite{Multimarket1525135} for the first two stages of the problem studied, namely the day-ahead and the automatic generation control (AGC) market. However, since the third-stage (balancing market) traded volumes are small, the producer can negatively affect the prices with its participation. Price scenarios are generated using ARIMA models for the two first stages, whereas for the third stage a linear curve with negative slope is used to represent the influence of the producer's offered capacity on the market price.
	
	Hydro-power plant participation in short-term markets accounting for the technical constraints and several reservoir levels is formulated and solved in \cite{Fleten2007916}. Optimal bidding curves for the participation of a ``price-taker" hydro-power producer in the Nordic spot market are derived accounting for price uncertainty. In \cite{Boomsma2014797}, the bidding strategy of a two-level reservoir plant is casted as a multi-stage stochastic program in order to represent the different sequential trading floors, namely the day-ahead spot market and the hour-ahead balancing market. The effects of coordinated bidding and the ``price-maker" versus ``price-taker" assumptions on the generated profits are evaluated. In \cite{PandZic2013}, bidding strategies for a virtual power plant (VPP) buying and selling energy in the day-ahead and the balancing market in the form of a multi-stage stochastic optimisation are investigated. The VPP aggregates a pumped hydro energy storage (PHES) unit as well as a conventional generator with stochastic intermittent power production and consumption. The goal of the VPP operator is the maximization of the expected profits under price uncertainty.
	
	In these approaches, the intraday market is considered as auction-based and it is modelled as a single recourse action. For each trading period, the optimal offered quantity is derived according to the realization of various stochastic variables. However, in reality, for most European countries, according to the EU Network Codes \cite{Meeus2017}, modern intraday market trading will primarily be a continuous process.
	
	The strategic participation in the CID market is investigated for the case of an RES producer in \cite{Henriot2014} and \cite{Garnier2014}. In both works, the problem is formulated as a sequential decision-making process, where the operator adjusts its offers during the trading horizon, according to the RES forecast updates for the physical delivery of power. Additionally, in \cite{Gonsch2016} the use of a PHES unit is proposed to undertake energy arbitrage and to offset potential deviations. The trading process is formulated as a Markov Decision Process (MDP) where the future commitment decision in the market is based on the stochastic realization of the intraday price, the imbalance penalty, the RES production and the storage availability.
	
	
	The volatility of the CID prices, along with the quality of the forecast updates, are found to be key factors that influence the degree of activity and success of the deployed bidding strategies \cite{Henriot2014}. Therefore, the CID prices and the forecast errors are considered as correlated stochastic processes in \cite{Garnier2014}. Alternatively, in \cite{Henriot2014}, the CID price is constructed as a linear function of the offered quantity with an increasing slope as the gate closure approaches. In this way, the scarcity of conventional units approaching real time is reflected. In \cite{Gonsch2016}, real weather data and market data are used to simulate the forecast error and CID price processes.
	
	For the sequential decision-making problem in the CID market, the offered quantity of energy is the decision variable to be optimised \cite{Garnier2014}. The optimisation is carried out using Approximate Dynamic Programming (ADP) methods, where a parameterised policy is obtained based on the observed stochastic processes for the price, the RES error and the level of the reservoir \cite{Gonsch2016}. The ADP approach presented in \cite{Gonsch2016} is compared in \cite{Hassler2017} to some threshold-based heuristic decision rules. The parameters are updated according to simulation-based experience and the obtained performance is comparable to the ADP algorithm. The obtained decision rules are intuitively interpretable and are derived efficiently through simulation-based optimisation.

	The bidding strategy deployed by a storage device operator participating in a slightly different real-time market organized by NYISO is presented in \cite{Jiang2014}. In this market, the commitment decision is taken one hour ahead of real-time and the settlements occur intra-hour every five minutes. In this setting, the storage operator selects two price thresholds at which the intra-hour settlements occur. The problem is formulated as an MDP and is solved using an ADP algorithm that exploits a particular monotonicity property. A distribution-free variant that assumes no knowledge of the price distribution is proposed. The optimal policy is trained using historical real-time price data.

	Even though the focus of the mentioned articles lies on the CID market, the trading decisions are considered to take place in discrete time-steps. A different approach is presented in \cite{aid2016optimal}, where the CID market participation is modelled as a continuous time process using stochastic differential equations (SDE). The Hamilton Jacobi Bellman (HJB) equation is used for the determination of the optimal trading strategy. The goal is the minimization of the imbalance cost faced by a power producer arising from the residual error between the RES production and demand. The optimal trading rate is derived assuming a stochastic process for the market price using real market data and the residual error.
	

	
	In the approaches presented so far, the CID price is modelled as a stochastic process assuming that the participating agent is a ``price-taker". However, in the CID market, this assumption implies that the CID market is liquid and the price at which one can buy or sell energy at a given time are similar or the same. This assumption does not always hold, since the mean bid-ask spread in a trading session in the German intraday market for 2015 was several hundred times larger than the tick-size (i.e. the minimum price movement of a trading instrument) \cite{Balardy2018}. It is also reported in the same study that the spread decreases as trading approaches the gate closure. 
	
	An approach that explicitly considers the order book is presented in \cite{Bertrand2019}. A threshold-based policy is used to optimise the bid acceptance for storage units participating in the CID market. A collection of different factors such as the time of the day are used for the adaptation of the price thresholds. The threshold policy is trained using a policy gradient method (REINFORCE) and the results show improved performance against the ``rolling intrinsic" benchmark. In this paper, we present a methodology that serves as a wrapper around the existing industrial state of the art. We provide a generic modelling framework for the problem and we elaborate on all the assumptions that allow the formulation of the problem as an MDP. We solve the resulting problem using a value function approximation method.

	\subsection{Contribution of the paper}
	In this paper, we focus on the sequential decision-making problem of a storage device operator participating in the CID market. Firstly, we present a novel modelling framework for the CID market, where the trading agents exchange energy via a centralized order book. Each trading agent is assumed to dynamically select the orders that maximize its benefits throughout the trading horizon. In contrast to the existing literature, all the available orders are considered explicitly with the intention to uncover more information at each trading decision. The liquidity of the market and the influence of each agent on the price are directly reflected in the observed order book, and the developed strategies can adapt accordingly. Secondly, we model explicitly the dynamics of the storage system. Finally, the resulting problem is cast as an MDP.
	
	The intraday trading problem of a storage device is solved using Deep Reinforcement Learning techniques, specifically an asynchronous distributed variant of the fitted Q iteration RL algorithm with deep neural networks as function approximators \cite{ernst2005tree}. Due to the high-dimensionality and the dynamically evolving size of the order book, we motivate the use of high-level actions and a constant size, low-dimensional state representation. The agent can select between trading and idling. The goal of the selected actions is the identification of the opportunity cost of trading given the state of the order book observed and the maximization of the total return over the trading horizon. The resulting optimal policy is evaluated using real data from the German ID market \cite{EPEX}. In summary, the contributions of this work are the following:
	
	\begin{itemize}
		\item We model the CID market trading process as an MDP where the energy exchanges occur explicitly through a centralized order book. The trading policy is adapted to the state of the order book. 
		\item The operational constraints of the storage device are considered explicitly.
		\item A state space reduction is proposed in order to deal with the size of the order book.
		\item A novel representation of high-level actions is used to identify the opportunity cost between trading and idling.
		\item The fitted Q iteration algorithm is used to find a time-variant policy that maximizes the total cumulative profits collected.
		\item Artificial trajectories are produced using historical data from the German CID market in order to address exploration issues.
	\end{itemize}
	
	\subsection{Outline of the paper}
	The rest of the paper is organized as follows. In Section \ref{sec: BiddingProcess}, the CID market trading framework is presented. The interaction of the trading agents via a centralized order book is formulated as a dynamic process. All the available information for an asset trading agent is detailed and the objective is defined as the cumulative profits. In Section \ref{sec: RLProblem}, all the assumptions necessary to formulate the bidding process in the CID market as an MDP are listed. The methodology utilised to find an optimal policy that maximizes the cumulative profits of the proposed MDP is detailed in Section \ref{sec: Methodology}. A case study using real data from the German CID market is performed in Section \ref{sec: Results}. The results as well as considerations about limitations of the developed methodology are discussed in Section \ref{sec: Discussions}. Finally, conclusions of this work are drawn and future recommendations are provided in Section \ref{sec: Conclusions}. A detailed nomenclature is provided at the Appendix \ref{sec: Appendix}.
	
	\section{Continuous Intraday Bidding process} \label{sec: BiddingProcess}
	\begin{figure*}
		\centering
		\begin{tikzpicture}[scale=0.8 , spy using outlines={circle, magnification=4, size=5cm, connect spies}]
		\spy [black] on (2.85,-1.6) in node [above] at (2,1);
		\draw (0,0) -- (1,0);
		\draw decorate [decoration={snake}] {(1,0) -- (3.5,0)};
		\draw (3.5,0) -- (12,0);
		\draw (10,2) -- (18,2);
		\draw (0,-2) -- (12,-2);

		\foreach \x in {10,12,14,16,18}
		\draw (\x cm,2.2) -- (\x cm,1.8);
		\foreach \x in {0,6,8,10,12}
		\draw (\x cm,6pt) -- (\x cm,-6pt);
		\foreach \x in {0,6,8,10,12}
		\draw (\x cm,-2.2) -- (\x cm,-1.8);
		\foreach \x in {0,0.5,1,1.5,2,2.5,3,3.5,4,4.5,5,5.5,6.5,7,7.5,8.5,9,9.5,10.5,11,11.5,12}
		\draw (\x cm,-2.1) -- (\x cm,-1.9);
		
		\draw[<->,thin] (10,2.1) -- (12,2.1)
		node[pos=0.5,above]{\footnotesize{$\lambda(Q_1)$}};
		
		\draw[->,thin] (8,-3) -- (10,-3)
		node[pos=0.5,below]{$t$};
		
		\draw[->,thin] (13.2,1) -- (14.8,1)
		node[pos=0.5,below]{$\tau$};
		
		
		\draw (0,0) node[below=6pt] {16:00} node[above=6pt] {$ t^{Q_1}_{open},t^{Q_2}_{open},t^{Q_3}_{open},t^{Q_4}_{open} $};
		\draw (6,0) node[below=6pt] { 23:30 } node[above=6pt] {$ t^{Q_1}_{close} $};
		\draw (8,0) node[below=6pt] { 23:45} node[above=6pt] {$ t^{Q_2}_{close} $};
		\draw (10,0) node[below=6pt] { 00:00 } node[above=6pt] {$ t^{Q_3}_{close} $};
		\draw (12,0) node[below=6pt] { 00:15 } node[above=6pt] {$ t^{Q_4}_{close} $};
		
		\draw (0,-2) node[below=6pt] {16:00} node[above=6pt] {$ t^{Q_1}_{open},t^{Q_2}_{open},t^{Q_3}_{open},t^{Q_4}_{open} $};
		\draw (6,-2) node[below=6pt] { 23:30 } node[above=6pt] {$ t^{Q_1}_{close} $};
		\draw (8,-2) node[below=6pt] { 23:45} node[above=6pt] {$ t^{Q_2}_{close} $};
		\draw (10,-2) node[below=6pt] { 00:00 } node[above=6pt] {$ t^{Q_3}_{close} $};
		\draw (12,-2) node[below=6pt] { 00:15 } node[above=6pt] {$ t^{Q_4}_{close} $};

		\draw (3,-2) node[below=1pt] { {\fontsize{3pt}{48} \selectfont \textcolor{black}{$t-1$}}} node[above=2pt] {{\fontsize{3pt}{48} \selectfont \textcolor{black}{$ a_{i,t-1} $}} };
		\draw (3.5,-2) node[below=1pt] { {\fontsize{3pt}{48} \selectfont \textcolor{black}{$t$}}} node[above=2pt] {{\fontsize{3pt}{48} \selectfont \textcolor{black}{$a_{i,t} $}} };
		\draw (4,-2) node[below=1pt] { {\fontsize{3pt}{48} \selectfont \textcolor{black}{$t+1$}}} node[above=2pt] {{\fontsize{3pt}{48} \selectfont \textcolor{black}{$ a_{i,t+1} $}} };
		
		\draw[<->,ultra thin,red] (3,-1.95) -- (3.5,-1.95)
		node[pos=0.5,above=-2pt]{{\fontsize{3pt}{48} \selectfont \textcolor{red}{$a_{-i,t}$}}};
		\draw[<->,ultra thin,red] (3.5,-1.95) -- (4,-1.95)
		node[pos=0.5,above=-2pt]{{\fontsize{3pt}{48} \selectfont \textcolor{red}{$a_{-i,t+1}$}}};

		\draw (6.5,2) node[] {Delivery timeline $\bar{T}$};
		\draw (16,0) node[] {Continuous trading timeline};
		\draw (16,-2) node[] {Discrete trading timeline $T$};

		\draw (10,2) node[below=6pt] { 00:00 } node[above=6pt] {$ t^{Q_1}_{delivery}$};
		\draw (12,2) node[below=6pt] { 00:15 } node[above=6pt] {$ t^{Q_2}_{delivery} $};
		\draw (12,2) node[above=24pt] {$ t^{Q_1}_{settle} ,$};
		\draw (14,2) node[below=6pt] { 00:30 } node[above=6pt] {$ t^{Q_3}_{delivery} $};
		\draw (14,2) node[above=24pt] {$ t^{Q_2}_{settle} ,$};
		\draw (16,2) node[below=6pt] { 00:45 } node[above=6pt] {$ t^{Q_4}_{delivery} $};
		\draw (16,2) node[above=24pt] {$ t^{Q_3}_{settle} ,$};
		\draw (18,2) node[below=6pt] { 01:00 } node[above=6pt] {$ $};
		\draw (18,2) node[above=24pt] {$ t^{Q_4}_{settle} $};
		
		\end{tikzpicture}
		\caption{Trading (continuous and discrete) and delivery timelines for products $Q_1$ to $Q_4$}\label{fig: Timeline}
	\end{figure*}
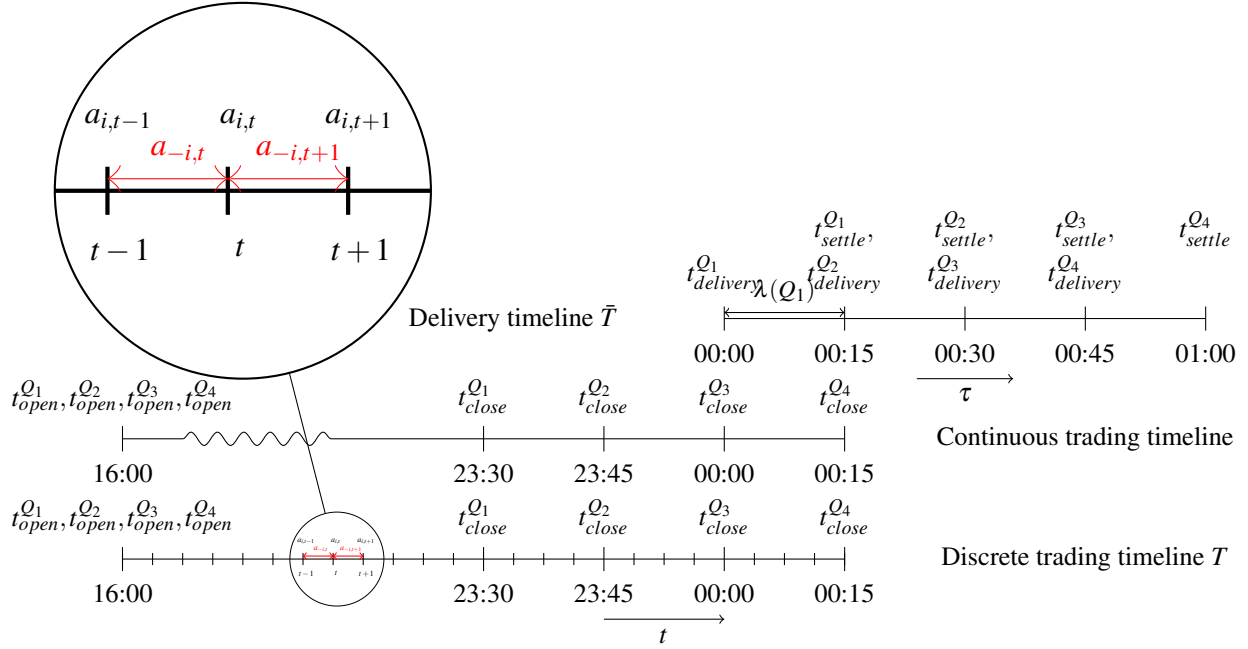

	\subsection{Continuous Intraday market design} \label{sec: CIDDesign}
	
	The participation in the CID market is a continuous process similar to the stock exchange. Each market product $x \in X $, where $ X $ is the set of all available products, is defined as the physical delivery of energy in a pre-defined time slot. The time slot corresponding to product $x$ is defined by its starting point $ t_{delivery}(x)$ and its duration $ \lambda(x) $. The trading process for time slot $x$ opens at $ t_{open}(x)$ and closes at $ t_{close}(x)$. During the time interval $t \in \left[ t_{open}(x),t_{close}(x)\right] $, a participant can exchange energy with other participants for the lagged physical delivery during the interval \timespan{x}, with: $$\timespan{x}=\left[ t_{delivery}(x),t_{delivery}(x)+\lambda(x)\right].$$ The exchange of energy takes place through a centralized order book that contains all the unmatched orders $o_{j}$, where $ j\in N_t$ corresponds to a unique index that every order receives upon arrival. The set $N_t \subseteq \mathbb{N}$ gathers all the unique indices of the orders available at time $t$. We denote the status of the order book at time $t$ by $ O_{t} = (o_{j},\forall j\in N_t)$. As time progresses new orders appear and existing ones are either accepted or cancelled. 
	
	Trading for a set of products is considered to start at the gate opening of the first product and to finish at the gate closure of the last product. More formally, considering an ordered set of available products $X = \{Q_1,...Q_{96}\}$, the corresponding trading horizon is defined as $T = \left[t_{open}(Q_1),t_{close}(Q_{96}) \right]$. For instance, in the German CID market, trading of hourly (quarterly) products for day $D$ opens at 3 pm (4 pm) of day $D-1$ respectively. For each product $x$, the gate closes 30 minutes before the actual energy delivery at $ t_{delivery}(x)$. The timeline for trading products $Q_1$ to $Q_4$ that correspond to the physical delivery in 15-minute time slots from 00:00 until 01:00, is presented in Figure \ref{fig: Timeline}. It can be observed that the agent can trade for all products until 23:30. After each subsequent gate closure the number of available products decreases and the commitment for the corresponding time slot is defined. Potential deviations during the physical delivery of energy are penalized in the imbalance market. 
	
	\subsection{Continuous Intraday market environment} \label{sec: CIDEnv}
	
	As its name indicates, the CID market is a continuous environment. In order to solve the trading problem presented in this paper, it has been decided to perform a relevant discretisation operation. As shown in Figure 1, the trading timeline is discretised in a high number of time-steps of constant duration $\Delta t$. Each discretised trading interval for product $x$ can be denoted by the set of time-steps $T(x) = \left\{t_{open}(x), t_{open}(x) + \Delta t, ..., t_{close}(x) - \Delta t , t_{close}(x)\right\}$. Then, the discrete-time trading opportunities for the entire set of products $X$ can be modelled such that the time-steps are defined as $t \in T= \bigcup_{x \in X} T(x)$. In the following, for the sake of clarity, the increment (decrement) operation $t+1$ ($t-1$) will be used to model the discrete transition from time-step $t$ to time-step $t + \Delta t$ ($ t - \Delta t$).
	
	It is important to note that in theory the discretisation operation leads to suboptimalities in the decision-making process. However, as the discretisation becomes finer ($\Delta t \rightarrow 0$), the decisions taken can be considered near-optimal. Increasing the granularity of the decision time-line results in an increase of the number of decisions that can be taken and hence, the size of the decision-making problem. Thus, there is a clear trade-off between complexity and quality of the resulting decisions when using a finite discretisation.
	
	Let $X_t$ denote the set of available products at time-step $t \in T$ such that:
	\begin{gather}
	X_t = \left\lbrace x | x\in X , t \leq t_{close}(x)\right\rbrace .\nonumber
	\end{gather}
	We define the state of the CID market environment at time-step $t$ as $s^{OB}_t = O_{t} \in S^{OB}$. The state contains the observation of the order book at time-step $t \in T$ i.e. the unmatched orders for all the available products $x\in X_t \subset X$.
	
	A set of $n$ agents $I=\{1,2,...,n\}$ are continuously interacting in the CID environment exchanging energy. Each agent $i \in I$ can express its willingness to buy or sell energy by posting at instant $t$ a set of new orders $a_{i,t} \in A_{i}$ in the order book, which results in the joint action $a_t = (a_{1,t},...,a_{n,t}) \in \prod_{i=1}^{n} A_{i}$. 
	
	The process of designing the set of new orders $a_{i,t}$ for agent $i$ at instant $t$ consists, for each new order, in determining the product $x \in X_t$, the side of the order $y \in \{``Sell",``Buy"\}$, the volume $v \in \mathbb{R}^{+}$, the price level $p \in \left[ p_{min}, p_{max}\right]$ of each unit offered to be produced or consumed, and the various validity and execution specifications $e \in E$. The index of each new order $j$ belongs to the set $j \in N'_t$.
	
	The set of new orders is defined as $a_{i,t} = ((x_j,y_j,v_j,p_j,e_j), \forall j \in N'_t \subseteq \mathbb{N})$. We will use the notation for the joint action $a_t = (a_{i,t},a_{-i,t})$ to refer to the action that agent $i$ selects $a_{i,t}$ and the joint action that all other agents use $a_{-i,t} = (a_{1,t},...,a_{i-1,t},a_{i+1.t},...,a_{n,t})$.
	
	\begin{table}[h]
		\centering
		\caption{Order Book for $Q_1$ and time slot 00:00-00:15}
		\label{table: Order Book}
		\begin{tabular}{lllllll}
			\textbf{$i$} & \textbf{Side} & \textbf{$v$ [MW]} & \textbf{$p$ [\euro/MWh]} & \\ \cmidrule[1pt]{1-4} 
			4 & ``Sell'' & 6.25 & 36.3 & \\
			2 & ``Sell'' & 2.35 & 34.5 & $\longleftarrow$ ask\\ \cmidrule{1-4} \morecmidrules \cmidrule{1-4}
			1 & ``Buy'' & 3.15 & 33.8 & $\longleftarrow$ bid\\
			3 & ``Buy'' & 1.125 & 29.3 & \\
			5 & ``Buy'' & 2.5 & 15.9 & 
		\end{tabular}
	\end{table}
	
	The orders are treated according to the first come first served (FCFS) rule. Table \ref{table: Order Book} presents an observation of the order book for product $Q_1$. The difference between the most expensive ``Buy" order (``bid") and the cheapest ``Sell" order (``ask") defines the bid-ask spread of the product. A deal between two counter-parties is struck when the price $p_{buy}$ of a ``Buy'' order and the price $p_{sell}$ of a ``Sell'' order satisfy the condition $p_{buy} \ge p_{sell}$. This condition is tested at the arrival of each new order. The volume of the transaction is defined as the minimum quantity between the ``Buy'' and ``Sell'' order ($\min(v_{buy},v_{sell})$). The residual volume remains available in the market at the same price. As mentioned in the previous section, each transaction is settled following the pay-as-bid principle, at the price indicated by the oldest order.
	
	Finally, at each time-step $t$, every agent $i$ observes the state of the order book $s_{t}^{OB}$, performs certain actions (posting a set of new orders) $a_{i, t}$, inducing a transition which can be represented by the following equation:
	
	\begin{equation}
	s_{t+1}^{OB} = f(s_{t}^{OB}, a_{i, t}, a_{-i, t}). \label{eqn: Transition}
	\end{equation}

	\subsection{Asset trading} \label{sec: Asset trading}

	An asset optimizing agent participating in the CID market can adjust its position for product $x$ until the corresponding gate closure $t_{close}(x)$. However, the physical delivery of power is decided at $t_{delivery}(x)$. An additional amount of information (potentially valuable for certain players) is received during the period $\left\lbrace t_{close}(x),..,t_{delivery}(x)\right\rbrace $, from the gate closure until the delivery of power. Based on this updated information, an asset-trading agent may need to or have an incentive to deviate from the net contracted power in the market. 
	
	Let $v_{i,t}^{con}= (v_{i,t}^{con}(x), \forall x \in X_t) \in \mathbb{R}^{|X_t|}$, gather the volumes of power contracted by agent $i$ for the available products $x \in X_t $ at each time-step $t\in T$. In the following, we will adopt the convention for $v_{i,t}^{con}(x)$ to be positive when agent $i$ contracts the net volume to sell (produce) and negative when the agent contracts the volume to buy (consume) energy for product $x$ at time-step $t$.
	
	Following each market transition as indicated by equation (\ref{eqn: Transition}), the volumes contracted $v_{i,t}^{con}$ are determined based on the transactions that have occurred. The contracted volumes $v_{i,t}^{con}$ are derived according to the FCFS rule that is detailed in \cite{EPEXRules}. The mathematical formulation of the clearing algorithm is provided in \cite{Le2019}. The objective function of the clearing algorithm is comprised of two terms, namely the social welfare and a penalty term modelling the price-time priority rule. The orders that maximize this objective are matched, provided that they satisfy the balancing equations and constraints related to their specifications. The clearing rule is implicitly given by:
	\begin{gather}
	v_{i,t}^{con} = clear(i, s^{OB}_t , a_{i,t}, a_{-i,t} ).\label{eqn: Vcontracted}
	\end{gather}
	
	We denote as $P_{i,t}^{mar}(x) \in \mathbb{R}$ the net contracted power in the market by agent $i$ for each product $x \in X$, which is updated at every time-step $t \in T$ according to: 
	\begin{gather}
	P_{i,t+1}^{mar}(x) = P_{i,t}^{mar}(x) + v_{i,t}^{con}(x).\label{eqn: Pmarket}\\
	\forall x \in X_t \nonumber
	\end{gather}
	
	The discretisation of the delivery timeline $\bar{T}$ is done with time-steps of duration $\Delta \tau$, equal to the minimum duration of delivery for the products considered. The discrete delivery timeline $\bar{T}$ is considered to start at the beginning of delivery of the first product $\tau_{init}$ and to finish at the end of the delivery of the last product $\tau_{term}$. For the simple case where only four quarterly products are considered, as shown in Figure \ref{fig: Timeline}, the delivery time-step is $\Delta \tau = 15 min$ and the delivery timeline $\bar{T} = \left\lbrace 00:00,00:15,...,01:00\right\rbrace $, where $\tau_{init} = 00:00$ and $\tau_{term}=01:00$. In general, when only one type of product is considered (e.g. quarterly), there is a straightforward relation between time of delivery $\tau$ and product $x$, since $\tau = t_{delivery}(x)$ and $\Delta \tau= \lambda(x)$. Thus, terms $x$ or $\tau$ can be used interchangeably. For the sake of keeping the notation relatively simple, we will only consider quarterly products in the rest of the paper. In such a context, the terms $P_{i,t}^{mar}(\tau)$ or $P_{i,t}^{mar}(x)$ can be used interchangeably to denote the net contracted power in the market by agent $i$ at trading step $t$ for delivery time-step $\tau$ (product $x$). 
	
	As the trading process evolves the set of delivery time-steps $\tau$ for which the asset-optimizing can make decisions decreases as trading time $t$ crosses the delivery time $\tau$. Let $\bar{T}(t) \subseteq \bar{T}$ be a function that yields the subset of delivery time-steps $\tau \in \bar{T}$ that follow time-step $t \in T$ such that:
	\begin{gather}
	\bar{T}(t) = \left\lbrace \tau | \tau\in \bar{T} \setminus \left\lbrace \tau_{term} \right\rbrace , t \leq \tau \right\rbrace. \nonumber
	\end{gather}
	
	The participation of an asset-optimizing agent in the CID market is composed of two coupled decision processes with different timescales. First, the trading process where a decision is taken at each time-step $t$ about the energy contracted until the gate closure $t_{close}(x)$. During this process, the agent can decide about its position in the market and create scenarios/make projections about the actual delivery plan based on its position. Second, the physical delivery decision that is taken at the time of the delivery $\tau$ or $t_{delivery}(x)$ based on the total net contracted power in the market during the trading process.
	
	An agent $i$ participating in the CID market is assumed to monitor the state of the order book $s_t^{OB}$ and its net contracted power in the market $P_{i, t}^{mar}(x)$ for each product $x \in X$, which becomes fixed once the gate closure occurs at $t_{close}(x)$. Depending on the role it presumes in the market, an asset-optimizing agent is assumed to monitor all the available information about its assets. We distinguish the three following cases among the many different roles that can be played by an agent in the CID market:
	\begin{itemize}
		
		\item \textit{The agent controls a physical asset that can generate and/or consume electricity}. We define as $G_{i,t}(\tau) \in \left[ \underline{G_{i}}, \overline{G_{i}}\right]$ the power production level for agent $i$ at delivery time-step $\tau$ as computed at trading step $t$. In a similar way, we define the power consumption level $C_{i,t}(\tau) \in \left[ \underline{C_{i}}, \overline{C_{i}} \right]$, where $\underline{C_{i}}, \overline{C_{i}},\underline{G_{i}}, \overline{G_{i}}\in \mathbb{R^{+}}$. We further assume that the actual production $g_{i,t}(t')$ and consumption level $c_{i,t}(t')$ during the time-period of delivery $t'\in \left[\tau, \tau+\Delta \tau\right)$, is constant for each product $x$ such that:
		\begin{gather}
		g_{i,t}(t')=G_{i,t}(\tau),\\
		c_{i,t}(t')=C_{i,t}(\tau),\\
		\forall t'\in \left[\tau, \tau+\Delta \tau\right). \nonumber
		\end{gather}
		
		At each time-step $t$ during the trading process, agent $i$ can decide to adjust its generation level by $\Delta G_{i,t}(\tau)$ or its consumption level by $\Delta C_{i,t}(\tau)$. According to these adjustments the generation and consumption levels can be updated at each time-step $t$ according to:
		\begin{gather}
		G_{i,t+1}(\tau) = G_{i,t}(\tau) + \Delta G_{i,t}(\tau), \label{eqn : Pgen}\\
		C_{i,t+1}(\tau) = C_{i,t}(\tau) + \Delta C_{i,t}(\tau), \label{eqn : Pcon}\\
		\forall \tau \in \bar{T}(t). \nonumber
		\end{gather}
		
		Let $w^{exog}_{i,t}$ denote any other relevant exogenous information to agent $i$ such as the RES forecast, a forecast of the actions of other agents, or the imbalance prices. The computation of $\Delta G_{i,t}(\cdot)$ and $\Delta C_{i,t}(\cdot)$ depends on the market position, the technical limits of the assets, the state of the order book and the exogenous information $w^{exog}_{i,t}$. We define the residual production $P^{res}_{i,t}(\tau)\in \mathbb{R}$ at delivery time-step $\tau$ as the difference between the production and the consumption levels and can be computed by:
		\begin{gather}
		P^{res}_{i,t}(\tau) = G_{i,t}(\tau) - C_{i,t}(\tau).\label{eqn: Presidual}
		\end{gather}
		
		We note that the amount of residual production $P^{res}_{i,t}(\tau)$ aggregates the combined effects that $G_{i,t}(\tau)$ and $C_{i,t}(\tau)$ have on the revenues made by agent $i$ through interacting with the markets (intraday/imbalance). 
		
		The level of generation and consumption for a market period $\tau$ can be adjusted at any time-step $t$ before the physical delivery $\tau$, but it becomes binding when $t=\tau$. We denote as $\Delta_{i,t}(\tau)$ the deviation from the market position for each time-step $\tau$, as scheduled at time $t$, after having computed the variables $G_{i,t}(\tau)$ and $C_{i,t}(\tau)$, as follows:
		\begin{gather}
		P_{i,t}^{mar}(\tau) + \Delta_{i,t}(\tau) = P^{res}_{i,t}(\tau), \label{eqn: ContractedPowerBalance}\\
		\forall \tau \in \bar{T}(t).\nonumber
		\end{gather}
		
		The term $\Delta_{i,t}(\tau)$ represents the imbalance for market period $\tau$ as estimated at time $t$. This imbalance may evolve up to time $t=\tau$. We denote by $\Delta_i(\tau) = \Delta_{i,t=\tau}(\tau)$ the final imbalance for market period $\tau$.

		The power balance of equation (\ref{eqn: ContractedPowerBalance}) written for time-step $t+1$ is given by:
		\begin{gather}
		P_{i,t+1 }^{mar}(\tau) + \Delta_{i,t+1}(\tau) = G_{i,t+1}(\tau) - C_{i,t+1}(\tau) \label{eqn: PowerBalance_tplus1}\\
		\forall \tau \in \bar{T}(t+1).\nonumber
		\end{gather}
		
		It can be observed that by substituting equations (\ref{eqn: Pmarket}), (\ref{eqn : Pgen}) and (\ref{eqn : Pcon}) in equation (\ref{eqn: PowerBalance_tplus1}) we have:
		
		\begin{align}
		P_{i,t}^{mar}(\tau) + v_{i,t}^{con}(\tau) + \Delta_{i,t+1}(\tau) = & \nonumber\\ G_{i,t}(\tau) + \Delta G_{i,t}(\tau) - (C_{i,t}(\tau)& + \Delta C_{i,t}(\tau)) \label{eqn: PowerBalance_t_1_sub}\\
		\forall \tau \in \bar{T}(t).\nonumber
		\end{align}
		
		The combination of equations (\ref{eqn: Presidual}) and (\ref{eqn: ContractedPowerBalance}) with equation (\ref{eqn: PowerBalance_t_1_sub}) yields the update of the imbalance vector according to:
		\begin{gather}
		\Delta_{i,t+1}(\tau) = \Delta_{i,t}(\tau)+ \Delta G_{i,t}(\tau) - \Delta C_{i,t}(\tau) - v_{i,t}^{con}(\tau) \label{eqn: ImbalanceUpdate}\\
		\forall \tau \in \bar{T}(t).\nonumber
		\end{gather}
		
		\item \textit{The agent does not own any physical asset (market maker)}. It is equivalent to the first case with $\underline{C_{i}}= \overline{C_{i}}=\underline{G_{i}}= \overline{G_{i}}=0$. The net imbalance $\Delta_{i,t}(\tau)$ is updated at every time-step $t\in T$ according to:
		\begin{gather}
		P_{i,t}^{mar}(\tau) + \Delta_{i,t}(\tau) = 0 , \\
		\forall \tau \in \bar{T}(t) .\nonumber
		\end{gather}
		
		\item \textit{The agent controls a storage device that can produce, store and consume energy}. We can consider an agent controlling a storage device as an agent that controls generation and production assets with specific constraints on the generation and the consumption level related to the nature of the storage device. Following this argument, let $G_{i,t}(\tau)$ ($C_{i,t}(\tau)$) refer to the level of discharging (charging) of the storage device for delivery time-step $\tau$, updated at time $t$. Obviously, if $G_{i,t}(\tau) > 0$ ($C_{i,t}(\tau) > 0$), then we automatically have $C_{i,t}(\tau) = 0$ ($G_{i,t}(\tau) = 0$) since a battery cannot charge and discharge energy at the same time. In this case, agent $i$ can decide to adjust its discharging (charging) level by $\Delta G_{i,t}(\tau)$ ($\Delta C_{i,t}(\tau)$). Let $\soc{i}{t}(\tau)$ denote the state of charge of the storage unit at delivery time-step $\tau \in \bar{T}$ as it is computed at time-step $t$, where $\soc{i}{t}(\tau) \in \left[\socmin{i}, \socmax{i} \right]$. The evolution of the state of charge during the delivery timeline can be updated at decision time-step $t$ as:
		
		\begin{align}
		\soc{i}{t}(\tau+\Delta \tau) = \soc{i}{t}(\tau) +& \nonumber\\ \Delta \tau \cdot \Bigg(\eta C_{i,t}(\tau) -& \frac{G_{i,t}(\tau)}{\eta} \Bigg), \label{eqn: SoCupdates}\\
		\forall \tau \in \bar{T}(t).\nonumber
		\end{align}
		Parameter $\eta$ represents the charging and discharging efficiencies of the storage unit which, for simplicity, we assume are equal. We note that for batteries, charging and discharging efficiencies may be different and depend on the charging/discharging speeds. As can be observed from equation (\ref{eqn: SoCupdates}), time-coupling constraints are imposed on $C_{i,t}(\tau)$ and $G_{i,t}(\tau)$ in order to ensure that the amount of energy that can be discharged during some period already exists in the storage device. Additionally, constraints associated with the maximum charging power $\overline{C}_i$ and discharging power $\overline{G}_i$, as well as the maximum and minimum energy level ($\socmax{i}$, $\socmin{i}$) are considered in order to model the operation of the storage device.

		Equation (\ref{eqn: SoCupdates}) can be written for time-step $t+1$ as:
		\begin{align}
		\soc{i}{t+1}(\tau+\Delta \tau) = \soc{i}{t+1}(\tau) +& \nonumber\\ \Delta \tau \cdot \Bigg(\eta C_{i,t+1}(\tau)& - \frac{G_{i,t+1}(\tau)}{\eta} \Bigg), \label{eqn: SoCupdates_1}\\
		\forall \tau \in \bar{T}(t+1).\nonumber
		\end{align}
		
		Combining equations (\ref{eqn: SoCupdates}) and (\ref{eqn: SoCupdates_1}) we can derive the updated vector of the state of charge at time-step $t+1$ depending on the decided adjustments ($\Delta G_{i,t}(\tau),\Delta C_{i,t}(\tau)$) as:
		\begin{align}
		SoC_{i,t+1}(\tau+\Delta \tau) - SoC_{i,t+1}(\tau)=&\nonumber \\
		SoC_{i,t}(\tau+\Delta \tau) - &SoC_{i,t}(\tau)+ \nonumber \\ \Delta \tau \cdot (\eta \Delta C_{i,t}(\tau)& - \frac{\Delta G_{i,t}(\tau)}{\eta}), \label{eqn: SoCupdates_2}\\
		\forall \tau \in \bar{T}(t).\nonumber
		\end{align}
		
		The state of charge $\soc{i}{t}(\tau)$ at delivery time-step $\tau$ can be updated until $t=\tau$. Let us also observe that there is a bijection between $P_{i,t}^{res}(\tau)$ and the terms $C_{i,t}(\tau)$ and $G_{i,t}(\tau)$ or, in other words, determining $P_{i,t}^{res}$ is equivalent to determining $C_{i,t}(\tau)$ and $G_{i,t}(\tau)$ and vice versa. The deviation from the committed schedule $\Delta_{i,t+1}(\tau)$ at delivery time-step $\tau$ at each time-step $t+1$ can be computed by equation (\ref{eqn: ImbalanceUpdate}).
		
		All the new information arriving at time-step $t$ for an asset-optimizing agent $i$ (controlling a storage device) is gathered in variable:
		\begin{align*}
		s_{i,t} = (& s^{OB}_t,\\
		&(P_{i,t}^{mar}(\tau),\Delta_{i,t}(\tau),G_{i,t}(\tau),C_{i,t}(\tau),SoC_{i,t}(\tau),\forall \tau \in \bar{T}), \\ 
		&w^{exog}_{i,t}) \in S_{i}.
		\end{align*}
		
	\end{itemize}
	The control action applied by an asset-optimizing agent $i$ trading in the CID market at time-step $t$ consists of posting new orders in the CID market and adjusting its production/consumption level or equivalently its charging/discharging level for the case of the storage device. The control actions can be summarised in variable $u_{i,t} = (a_{i,t},(\Delta C_{i,t}(\tau), \Delta G_{i,t}(\tau), \forall \tau \in \bar{T}))$. 
	
	
	In this paper, we consider that the trading agent adopts a simple strategy for determining, at each time-step $t$, the variables $\Delta C_{i,t}(\tau)$, $\Delta G_{i,t}(\tau)$ once the trading actions $a_{i,t}$ have been selected. In this case, the decision regarding the trading actions $a_{i,t}$ fully defines action $u_{i,t}$ and thus the notation $u_{i,t}$ will not be further used. This strategy will be referred to in the rest of the paper as the ``default" strategy for managing the storage device. According to this strategy, the agent aims at minimizing any imbalances ($\Delta_{i,t+1}(\tau)$) and therefore we use the following decision rule: 
	\begin{gather}
	\left( \Delta C_{i,t}(\tau),\Delta G_{i,t}(\tau),\forall \tau \in \bar{T} \right) = \argmin \sum_{\tau \in \bar{T}} \mid \Delta_{i,t+1}(\tau)\mid,\nonumber\\
	\mbox{ s.t. (\ref{eqn: Vcontracted}), (\ref{eqn: Pmarket}), (\ref{eqn: Presidual}), (\ref{eqn: ContractedPowerBalance}), (\ref{eqn: ImbalanceUpdate}), (\ref{eqn: SoCupdates})}.
	\end{gather}
	
	One can easily see that from equation (\ref{eqn: PowerBalance_t_1_sub}) this decision rule is equivalent to imposing $P^{res}_{i,t+1}(\tau)$ as close as possible to $P^{mar}_{i,t+1}(\tau)$, given the operational constraints of the device. We will elaborate later in this paper on the fact that adopting such a strategy is not suboptimal in a context where the agent needs to be balanced for every market period while being an aggressor in the CID market. 
	
	For the sake of simplicity, we assume that the decision process of an asset-optimizing agent terminates at the gate closure $t_{close}(x)$ along with the trading process. Thus, the final residual production $P^{res}_{i}(\tau)$ for delivery time-step $\tau$ is given by $P^{res}_{i}(\tau)= P^{res}_{i,t=t_{close}(x)}(\tau)$. Similarly, the final imbalance is provided by $\Delta_{i}(\tau)= \Delta_{i,t=t_{close}(x)}(\tau)$.
	
	Although this approach can be used for the optimisation of a portfolio of assets, in this paper, the focus lies on the case where the agent is operating a storage device. We note that this case is particularly interesting in the context of energy transition, where storage devices are expected to play a key role in the energy market.

	\subsection{Trading rewards}\label{sec: Trading_rewards}
	
	The instantaneous reward signal collected after each transition for agent $i$ is given by: 
	\begin{gather}
	r_{i,t} = R_{i} \left(t,s_{i,t},a_{i,t} ,a_{-i,t} \right),\label{eqn: rewardfunction}
	\end{gather}
	\noindent where $R_{i} : T \times S_{i} \times A_{1} \times ... \times A_{n} \to \mathbb{R}$. 
	
	The reward function $R_{i}$ is composed of the following terms:
	\begin{enumerate}[i.]
		\item The trading revenues obtained from the matching process of orders at time-step $t$, given by $\rho$ where $\rho$ is a stationary function $\rho : S^{OB} \times A_{1} \times ... \times A_{n} \to \mathbb{R}$,
		\item The imbalance penalty for deviation $\Delta_{i}(\tau)$ from the market position for delivery time-step $\tau$ at the imbalance price $I(\tau)$. The imbalance settlement process for product $x\in X$ (delivery time-step $\tau$) takes place at the end of the physical delivery $t_{settle}(x)$ (i.e. at $\tau+\Delta \tau$), as presented in Figure \ref{fig: Timeline}. We define the imbalance settlement timeline $T^{Imb}$, as $T^{Imb}=\left\lbrace \tau+\Delta \tau, \forall \tau \in \bar{T}\right\rbrace $. The imbalance penalty\footnote{The imbalance price $I(\tau)$ is defined by a process that depends on a plethora of factors among which is the net system imbalance during delivery period $\tau$, defined by the imbalance volumes of all the market players ($\sum^{I} \Delta_{i}(\tau)$). For the sake of simplicity we will assume that it is randomly sampled from a known distribution over prices that is not conditioned on any variable.} is only applied when time instance $t$ is an element of the imbalance settlement timeline.
	\end{enumerate}
	
	
	The function $R_{i}$ is defined as:
	\begin{gather}
	\begin{split}
	R_{i} \left(t,s_{i,t},a_{i,t} ,a_{-i,t} \right) &=\\
	\rho\left(s^{OB}_{t},a_{i,t} ,a_{-i,t} \right)& + 
	\begin{cases}
	\Delta_{i}(\tau)\cdot I(\tau)&, \text{if } t \in T^{Imb},\\
	0&, \text{otherwise}\end{cases} \label{eqn: reward}.
	\end{split}
	\end{gather}
	
	%
	

	\subsection{Trading policy} \label{sec: Trading policy}
	All the relevant information that summarises the past and that can be used to optimise the market participation is assumed to be contained in the history vector $h_{i,t} = (s_{i,0},a_{i,0}, r_{i,0},...,s_{i,t-1},a_{i,t-1}, r_{i,t-1},s_{i,t}) \in H_{i}$. Trading agent $i$ is assumed to select its actions following a non-anticipative history-dependent policy $\pi_{i}(h_{i,t})\in \Pi$ from the set of all admissible policies $\Pi$, according to: $a_{i,t} \sim \pi_{i} (\cdot | h_{i,t})$.
	
	\subsection{Trading objective} \label{sec: Trading_objective}
	
	The return collected by agent $i$ in a single trajectory $ \zeta = (s_{i,0},a_{i,0},...,a_{i,K-1},s_{i,K})$ of $K-1$ time-steps, given an initial state $s_{i,0} = s_{i}\in S_{i}$, which is the sum of cumulated rewards over this trajectory is given by:
	\begin{gather}
	G^{ \zeta}(s_{i}) = \sum_{t=0}^{K-1} R_{i}\left(t,s_{i,t},a_{i,t},a_{-i,t} \right) | s_{i,0} = s_{i} .\label{eqn: AgentsReturns}
	\end{gather}
	
	The sum of returns collected by agent $i$, where each agent $i$ is following a randomized policy $\pi_{i} \in \Pi$ are consequently given by:
	\begin{gather}
	V^{\pi_{i}}(s_{i}) = \E_{a_{i,t} \sim \pi_{i}, a_{-i,t} \sim \pi_{-i}} \Bigg\{\sum_{t=0}^{K-1} R_{i}\left(t,s_{i,t},a_{i,t},a_{-i,t} \right) | s_{i,0} = s_{i}\Bigg\} .\label{eqn: AgentsExpectedReturns}
	\end{gather}
	
	The goal of the trading agent $i$ is to identify an optimal policy $\pi_{i}^{*} \in \Pi$ that maximizes the expected sum of rewards collected along a trajectory. An optimal policy is obtained by:
	\begin{gather} 
	\pi_{i}^{*} = \argmax_{\pi_{i} \in \Pi} V^{\pi_{i}}(s_{i}) .\label{eqn: AgentsOptimalPolicy}
	\end{gather}
	
	\section{Reinforcement Learning Formulation} \label{sec: RLProblem}
	
	In this section, we propose a series of assumptions that allow us to formulate the previously introduced problem of a storage device operator trading in the CID market using a reinforcement learning (RL) framework. Based on these assumptions, the decision-making problem is cast as an MDP; the action space is tailored in order to represent a particular market player and additional restrictions on the operation of the storage device are introduced.
	
	\subsection{Assumptions on the decision process}
	
	\begin{asu}[\emph{Behaviour of the other agents}]\label{asu: OtherAgents}
		The other agents $-i$ interact with the order book in between two discrete time-steps in such a way that agent $i$ can be considered independent when interacting with the CID market at each time-step $t$. Moreover, it is assumed that these actions $a_{-i, t}$ depend strictly on the history of order book states $s_{t-1}^{OB}$ and thus by extension on the history $h_{i, t-1}$ for every time-step $t$:
		
		\begin{equation}
		a_{-i, t} \sim P_{a_{-i, t}}(\cdot | h_{i,t-1}) .\label{eqn: AgentsActions}
		\end{equation}
	\end{asu}
	
	Assumption (\ref{asu: OtherAgents}) suggests that the agents engage in a way that is very similar to a board game like chess, for which player $i$ can make a move only after players $-i$ have made their move. This behaviour is illustrated in Figure \ref{fig: Timeline} (magnified area). Given this assumption, the notation $a_{-i, t}$ can also be seen as referring to actions selected during the interval $(t-\Delta t, t)$.
	
	\begin{asu}[\emph{Exogenous information}]\label{asu: ExogenousInfo}
		The exogenous information $w^{exog}_{i,t}$ is given by a stochastic model that depends solely on $k$ past values, where $0 < k \leq t$ and a random disturbance $e_{i,t}$ according to:
		\begin{gather}
		w^{exog}_{i,t} = b(w^{exog}_{i,t-1},...,w^{exog}_{i,t-k},e_t), \label{eqn: Exogenous}\\
		e_{i,t} \sim P_{e_{i,t}}(\cdot | h_{i,t}) .\label{eqn: Exogenous_noise}
		\end{gather}
	\end{asu}

	\begin{asu}[\emph{Strategy for storage control}]\label{asu: StorageControl}
		The control decisions related to the charging ($\Delta C_{i,t}(\tau)$) or discharging ($\Delta G_{i,t}(\tau)$) power to/from the storage device are made based on the ``default" strategy described in Section \ref{sec: Asset trading}.
	\end{asu}

	It can be observed that with such an assumption, the storage control decisions ($\Delta C_{i,t}(\tau)$ and $\Delta G_{i,t}(\tau)$) are obtained as a direct consequence of the trading decisions $a_{i,t}$. Indeed, after the trading decisions are submitted and the market position is updated, the storage control decisions are subsequently derived following the ``default" strategy. Assumption (\ref{asu: StorageControl}) results in reducing the dimensionality of the action space and consequently the complexity of the decision-making problem.
	
	Following Assumptions (\ref{asu: OtherAgents}), (\ref{asu: ExogenousInfo}) and (\ref{asu: StorageControl}), one can simply observe that the decision-making problem faced by an agent $i$ operating a storage device and trading energy in the CID market can be formalised as a fully observable finite-time MDP with the following characteristics:
	
	\begin{itemize}
		\item \textit{Discrete time-step} $t \in T$, where $T$ is the optimisation horizon.
		\item \textit{State space} $H_{i}$, where the state of the system $h_{i,t} \in H_{i}$ at time $t$ summarises all past information that is relevant for future optimisation.
		\item \textit{Action space} $A_{i}$, where $a_{i,t} \in A_{i}$ is the set of new orders posted by agent $i$ at time-step $t$.
		\item \textit{Transition probabilities} $h_{i,t+1} \sim P(\cdot|h_{i,t},a_{i,t})$, that can be inferred by the following processes:	
		\begin{enumerate}
			\item $a_{-i,t} \in A_{-i}$ is drawn according to equation (\ref{eqn: AgentsActions})
			\item The state of the order book $s_{t+1}^{OB}$ follows the transition given by equation (\ref{eqn: Transition})
			\item The exogenous information $w^{exog}_{i,t}$ is given by equation (\ref{eqn: Exogenous}) and the noise by (\ref{eqn: Exogenous_noise})
			\item The variable $s_{i,t+1}$ that summarises the information of the storage device optimizing agent follows the transition given by equations (\ref{eqn: Transition}), (\ref{eqn : Pgen})-(\ref{eqn: ImbalanceUpdate}) (\ref{eqn: Exogenous}), (\ref{eqn: Exogenous_noise}) and (\ref{eqn: SoCupdates_2})
			\item The instantaneous reward $r_{i,t}$ collected after each transition is given by equations (\ref{eqn: rewardfunction}) and (\ref{eqn: reward}).
		\end{enumerate}
		The elements resulting from these processes can be used to construct $h_{i,t+1}$ in a straightforward way.

	\end{itemize}
	
	\subsection{Assumptions on the trading actions} \label{sec: Action_Assumptions}
	\begin{asu}[\emph{Aggressor}]\label{asu: Aggressor}
		The trading agent can only submit new orders that match already existing orders at their price (i.e. aggressor or liquidity taker).
	\end{asu}
	
	Let $A_{i}^{red}$ be the space that contains only actions that match pre-existing orders in the order book. According to Assumption (\ref{asu: Aggressor}), the $i^{th}$ agent, at time-step $t$, is restricted to select actions $a_{i,t} \in A_{i}^{red} \subset A_{i}$. Let $s^{OB}_{t} = ((x^{OB}_j,y'^{OB}_j,v^{OB}_j,p^{OB}_j,e^{OB}_j), \forall j \in N_{t})$ be the order book observation at trading time-step $t$. We use $y'^{OB}$ to denote that the new orders have the opposite side (``Buy" or ``Sell") than the existing orders. We denote as $a_{i,t}^{j}\in [0,1]$ the fraction of the volume accepted from order $j$. The reduced action space $A_{i}^{red}$ is then defined as: 
	\begin{gather}
	A_{i}^{red} = \{ (x^{OB}_j,y'^{OB}_j, a_{i,t}^{j}\cdot v^{OB}_j,p^{OB}_j,e^{OB}_j), a_{i,t}^{j}\in [0,1] , \forall j \in N_{t}\}.\nonumber
	\end{gather}
	At this point, posting a new set of orders $a_{i,t} \in A_{i}^{red}$ boils down to simply specifying the vector of fractions: $$\bar{a}_{i,t} = \left( a_{i,t}^{j}, \forall j \in N_t\right)\in \bar{A}_{i}^{red}$$ that define the partial or full acceptance of the existing orders. The action $a_{i,t}$ submitted by an aggressor is a function $l$ of the observed order book $s^{OB}_{t}$ and the vector of fractions $\bar{a}_{i,t}$ and is given by:
	\begin{gather}
	a_{i,t}=l(s^{OB}_{t},\bar{a}_{i,t}) .\label{eqn: ReducedAction}
	\end{gather}
	
	\subsection{Restrictions on the storage operation} \label{sec: Operation_Assumptions}
	
	\begin{asu}[\emph{No imbalances permitted}]\label{asu: Zero_Imbalances}
		The trading agent can only accept an order to buy or sell energy if and only if it does not result in any imbalance for the remaining delivery periods.
	\end{asu}
	According to Assumption (\ref{asu: Zero_Imbalances}) the agent is completely risk-averse in the sense that, even if it stops trading at any given point, its position in the market can be covered without causing any imbalance. This assumption is quite restrictive with respect to the full potential of an asset-optimizing agent in the CID market. We note that, according to the German regulation policies (see \cite{Braun2016}), the imbalance market should not be considered as an optimisation floor and the storage device should always be balanced at each trading time-step $t$ ($\Delta_{i,t}(\tau)=0, \forall \tau \in \bar{T}$). In this respect, we can view Assumption \ref{asu: Zero_Imbalances} as a way to comply with the German regulation policies in a risk-free context where each new trade should not create an imbalance that would have to be covered later.
	
	\begin{asu}[\emph{Optimization decoupling}]\label{asu: DaysDecoupling}
		The storage device has a given initial value for the storage level $SoC_{i}^{init}$ at the beginning of the delivery timeline. Moreover, it is constrained to terminate at a given level $SoC_{i}^{term}$ at the end of the delivery timeline.
	\end{asu}
	
	Under Assumption (\ref{asu: DaysDecoupling}) the optimisation of the storage unit over a long trading horizon can be decomposed into shorter optimisation windows (e.g. of one day). In the simulation results reported later in this paper, we will choose $SoC_{i}^{init}=SoC_{i}^{term}$.

	\section{Methodology} \label{sec: Methodology}
	
	In this section, we describe the methodology that has been applied for tackling the MDP problem described in subsection \ref{sec: RLProblem}. We consider that, in reality, an asset-optimizing agent has at its disposal a set of trajectories (one per day) from participating in the CID market in the past years. The process of collecting these trajectories and their structure is presented in Section \ref{sec: History_trajectories}. Based on this dataset, we propose in subsection \ref{sec: Solution} the deployment of the fitted Q iteration algorithm as introduced in \cite{ernst2005tree}. This algorithm belongs to the class of batch-mode RL algorithms that make use of all the available samples at once for updating the policy. This class of algorithms is known to be very sample efficient. 
	
	Despite the different assumptions made on the operation of the storage device and the way it is restricted to interact with the market, the dimensionality of the action space still remains very high. Due to limitations related to the function approximation architecture used to implement the fitted Q iteration algorithm, a low-dimensional and discrete action space is necessary, as discussed in subsection \ref{sec: Limitations}. Therefore, as part of the methodology, in subsection \ref{sec: HLA} we propose a way for reducing the action space. Afterwards, in subsection \ref{sec: SSR}, a more compact representation of the state space is proposed in order to reduce the computational complexity of the training process and increase the sample efficiency of the algorithm. 
	
	Finally, the low number of available samples (one trajectory per day) gives rise to issues related to the limited exploration of the agent. In order to address these issues, we generate a large number of trading trajectories of our MDP according to an $\epsilon$-greedy policy, using historical trading data. In the last part of this section, we elaborate on the strategy that is used in this paper for generating the trajectories and the limitations of this procedure.

	\subsection{Collection of trajectories} \label{sec: History_trajectories}
	
	As previously mentioned, an asset-optimizing agent can collect a set of trajectories from previous interactions with the CID market. Based on Assumption (\ref{asu: DaysDecoupling}), each day can be optimised separately and thus, trading for one day corresponds to one trajectory. We consider that the trading horizon defined in Section \ref{sec: CIDEnv} consists of $K$ discrete trading time-steps such that $T= \left\lbrace 0,...,K\right\rbrace $. A single trajectory sampled from the MDP described in Section \ref{sec: RLProblem} is defined as: 
	\begin{gather}
	\zeta_{m} = \left( h_{i,0}^{m}, a_{i,0}^{m},r_{i,0}^{m}, ...,h_{i,K-1}^{m},a_{i,K-1}^{m},r_{i,K-1}^{m},h_{i,K}^{m}\right) .\nonumber
	\end{gather}
	
	A set of $M$ trajectories can be then defined as:
	\begin{gather}
	F= \left\lbrace \zeta_{m}, m=1,...,M \right\rbrace .\nonumber
	\end{gather}
	
	The set of trajectories $F$ can be used to generate the set of sampled one-step system transitions $F'$ defined as:
	
	\begin{gather}
	F'=
	\left\lbrace 
	\begin{matrix} 
	(h_{i,0}^{1}, a_{i,0}^{1},r_{i,0}^{1},h_{i,1}^{1}), &\cdots& (h_{i,K-1}^{1}, a_{i,K-1}^{1},r_{i,K-1}^{1},h_{i,K}^{1}), \\
	\vdots&\ddots&\vdots\\
	(h_{i,0}^{M}, a_{i,0}^{M},r_{i,0}^{M},h_{i,1}^{M}), &\cdots& (h_{i,K-1}^{M}, a_{i,K-1}^{M},r_{i,K-1}^{M},h_{i,K}^{M}) 
	\end{matrix} 
	\right\rbrace .
	\nonumber
	\end{gather}
	
	The set $F'$ is split into $K$ sets of one-step system transitions $F'_{t}$ defined as:
	\begin{gather}
	F'_{t}= \left\lbrace (h_{i,t}^{m}, a_{i,t}^{m},r_{i,t}^{m},h_{i,t+1}^{m}), m=1,...,M \right\rbrace_{t}, \nonumber\\
	\forall t \in \left\lbrace 0,...,K-1 \right\rbrace .\nonumber
	\end{gather}
	
	A batch-mode RL algorithm can be implemented for extracting a relevant trading policy from these one-step system transitions. In the following subsection, the type of RL algorithm used for inferring a high-quality policy from this set of one-step system transitions is explained in detail. 
	
	\subsection{Batch-mode reinforcement learning} \label{sec: Solution}
	
	\textbf{Q-functions and Dynamic Programming}: In this section, the fitted Q iteration algorithm is proposed for the optimisation of the MDP defined in Section \ref{sec: RLProblem}, using a set of collected trajectories. In order to solve the problem, we first define the $Q$-function for each state-action pair ($h_{i,t},a_{i,t} $) at time $t$ as proposed in \cite{bertsekas2005dynamic} as:
	
	\begin{gather}
	Q_t(h_{i,t},a_{i,t}) = \hspace{-7pt} \underset{\footnotesize{\begin{array}{c} a_{-i,t},\mbox{ } e_{i,t}\end{array}}} \E \hspace{-7pt} \left\lbrace r_{i,t} + V_{t+1}(h_{i,t+1}) \right\rbrace \label{eqn: Qvalue}, \\ 
	\forall t \in \left\lbrace 0,...,K-1\right\rbrace .\nonumber
	\end{gather}
	
	A time-variant policy $\pi = \left\lbrace \mu_{0},...,\mu_{K-1} \right\rbrace \in \Pi$, consists in a sequence of functions $\mu_{t}$, where $\mu_{t}: H_{i} \rightarrow A_{i}^{red}$. An action $a_{i,t}$ is selected from this policy at each time-step $t$, according to $a_{i,t} = \mu_{t}(h_{i,t})$. We denote as $\pi^{t+1} = \left\lbrace \mu_{t+1},...,\mu_{K-1} \right\rbrace$ the sequence of functions $\mu_{t}$ from time-step $t+1$ until the end of the horizon. Standard results from dynamic programming (DP) show that for the finite time MDP we are addressing in this paper, there exists at least one such time-variant policy which is an optimal policy as defined by equation (\ref{eqn: AgentsOptimalPolicy}). Therefore, we focus on the computation of such an optimal time-variant policy. We define the value function $V_{t+1}$ as the optimal expected cumulative rewards from stage $t+1$ until the end of the horizon $K$ given by:
	
	\begin{align}
	V_{t+1}(h_{i}) = &&\nonumber \\
	\max_{\pi^{t+1} \in \Pi} \hspace{-7pt} \underset{\tiny{\begin{array}{c} (a_{-i,t+1},e_{i,t+1}) \\ \cdots\\ (a_{-i,K-1},e_{i,K-1}) \end{array}}} \E \hspace{-7pt} &\left\lbrace \sum_{k = t+1}^{K-1} R_{i,k} \left(h_{i,k}, \mu_{k}(h_{i,k}), a_{-i,k} \right) | h_{i,t+1} = h_{i} \right\rbrace.&\label{eqn: ValueFunction}
	\end{align}
	
	We observe that $Q_t(h_{i,t},a_{i,t} )$ is the value attained by taking action $a_{i,t}$ at state $h_{i,t}$ and subsequently using an optimal policy. Using the dynamic programming algorithm \cite{bertsekas2005dynamic} we have:
	
	\begin{gather}
	V_{t}(h_{i,t}) = \max_{a_{i,t} \in A_{i}^{red}} Q_t(h_{i,t},a_{i,t}) .\label{eqn: OptValueFunction}
	\end{gather}
	
	Equation (\ref{eqn: Qvalue}) can be written in the following form that relates $Q_t$ and $Q_{t+1}$:
	
	\begin{gather}
	Q_t(h_{i,t},a_{i,t}) = \hspace{-7pt} \underset{\tiny{\begin{array}{c} a_{-i,t},\mbox{ } e_{i,t}\end{array}}} \E \hspace{-7pt} \left\lbrace r_{i,t} + \max_{a_{i,t+1} \in A_{i}^{red}} Q_{t+1}(h_{i,t+1},a_{i,t+1}) \right\rbrace .\label{eqn: DynamicProgramming}
	\end{gather}
	
	An optimal time-variant policy $\pi^{*} = \left\lbrace \mu^{*}_{0},...,\mu^{*}_{K-1} \right\rbrace$ can be identified using the $Q$-functions as following:
	
	\begin{gather}
	\mu^{*}_{t} = \argmax_{a_{i,t} \in A_{i}^{red}} Q_t(h_{i,t},a_{i,t}), \label{eqn: OptPolicy}\\
	\forall t \in \left\lbrace 0,...,K-1\right\rbrace .\nonumber
	\end{gather}
	
	\textbf{Computing the Q-functions from a set of one-step system transitions}:
	In order to obtain the optimal time-variant policy $\pi^{*}$, the effort is focused on computing the Q-functions defined in equation (\ref{eqn: DynamicProgramming}). However, two aspects render the use of the standard value iteration algorithm impossible for solving the MDP defined in Section \ref{sec: RLProblem}. First, the transition probabilities of the MDP defined in Section \ref{sec: RLProblem} are not known. Instead, we can exploit the set of collected historical trajectories to compute the exact Q-functions using an algorithm such as Q-learning (presented in \cite{watkins1992}). Q-learning is designed for working only with trajectories, without any knowledge of the transition probabilities. Optimality is guaranteed given that all state-action pairs are observed infinitely often within the set of the historical trajectories and that the successor states are independently sampled at each occurrence of a state-action pair \cite{bertsekas2005dynamic}. In Section \ref{sec: Trajectories_Generation} we discuss the validity of this condition and we address the problem of limited exploration by generating additional artificial trajectories. Second, due to the continuous nature of the state and action spaces a tabular representation of the Q-functions used in Q-learning is not feasible. In order to overcome this issue, we use a function approximation architecture to represent the Q-functions \cite{busoniu2017}. 
	
	The computation of the approximate Q-functions is performed using the fitted Q iteration algorithm \cite{ernst2005tree}. We present the algorithm for the case where a parametric function approximation architecture ($Q_{t} (h_{i,t},a_{t};\theta_{t})$) is used (e.g. neural networks). In this case, the algorithm is used to compute, recursively, the parameter vectors $\theta_t$ starting from $t =K-1$. However, it should be emphasized that the fitted Q iteration algorithm can be adapted in a straightforward way to the case in which a non-parametric function approximation architecture is selected. 
	
	The set of $M$ samples of quadruples $F'_t = \left\lbrace (h_{i,t}^{m},a_{i,t}^{m},r_t^{m},h_{i,t+1}^{m}) , m=1,...,M \right\rbrace $ obtained from previous experience is exploited in order to update the parameter vectors $\theta_t$ by solving the supervised learning problem presented in equation (\ref{eqn: LS}). The target vectors $y_t$ are computed using the $Q$-function approximation of the next stage ($Q_{t+1} (h_{i,t+1},a_{t+1};\theta_{t+1})$) according to equation (\ref{eqn: targets}). The $Q$-function for the terminal state is set to zero ($\hat{Q}_K \equiv 0$) and the algorithm iterates backwards in the time horizon $T$, producing a sequence of approximate $Q$-functions denoted by $\hat{Q}= \{\hat{Q}_0,...,\hat{Q}_{K-1}\}$ until termination at $t = 0$.
	
	\begin{gather}
	\theta_t = \argmin_{\theta_t} \sum_{m=1}^{M} (Q_t(h^{m}_{i,t},a^{m}_t;\theta_t) - y^{m}_t)^2\label{eqn: LS}\\
	y^{m}_t = r^{m}_t + \max_{a_{i,t+1} \in A_{i}^{red}} Q_{t+1}(h^{m}_{i,t+1},a_{i,t+1};\theta_{t+1})\label{eqn: targets}
	\end{gather}
	Once the parameters $\theta_t$ are computed, the time-variant policy $\hat{\pi}^{*} = \left\lbrace \hat{\mu}^{*}_{0},...,\hat{\mu}^{*}_{K-1} \right\rbrace$ is obtained as:
	\begin{gather}
	\hat{\mu}^{*}_{t}(h_{i,t}) = \argmax_{a_{i,t} \in A_{i}^{red}} Q_t(h_{i,t},a_{i,t};\theta_t), \label{eqn: OptAppPolicy}\\
	\forall t \in \left\lbrace 0,...,K-1\right\rbrace .\nonumber
	\end{gather}

	In practice, a new trajectory is collected after each trading day. The set of collected trajectories $F$ is consequently augmented. Thus, the fitted Q iteration algorithm can be used to compute a new optimal policy when new data arrive. 
	
	\subsection{Limitations} \label{sec: Limitations}
	
	The fitted Q iteration algorithm, described in the previous section, can be used to provide a trading policy based on the set of past trajectories at the disposal of the agent. Even though, this approach is theoretically sound, in practice there are several limitations to overcome. The efficiency of the described fitted Q iteration algorithm is overshadowed by the high-dimensionality of the state and the action space. 
	
	The state variable $$h_{i,t} = (s_{i,0},a_{i,0}, r_{i,0},...,s_{i,t-1},a_{i,t-1}, r_{i,t-1},s_{i,t}) \in H_{i}$$ is composed of :
	\begin{itemize}
		\item The entire history of actions $\left( a_{i,0},...,a_{i,t-1}\right)$ before time $t$
		\item The entire history of rewards $\left( r_{i,0},...,r_{i,t-1}\right)$ before time $t$
		\item The history of order book states $\left( s_{0}^{OB},...,s_{t}^{OB}\right) $ up to time $t$ and,y of the private information $( s_{i,0}^{private},...,s^{private}_{i,t}) $ up to time $t$, where:
		\begin{align}
		s^{private}_{i,t} = &((P_{i,t}^{mar}(\tau),\Delta_{i,t}(\tau),\nonumber\\
		&G_{i,t}(\tau),C_{i,t}(\tau),SoC_{i,t}(\tau),\forall \tau \in \bar{T}),\nonumber\\
		&w^{exog}_{i,t}).\nonumber
		\end{align}
	\end{itemize}
	
	The state space $H_{i}$ as well as the action space $ A_{i}^{red}$, as described in Section \ref{sec: Action_Assumptions}, depend explicitly on the content of the order book $s^{OB}_{t}$. The dimension of these spaces at each time-step $t$ depends on the total number of available orders $\mid N_{t} \mid$ in the order book. However, the total number of orders is changing at each step $t$. Thus, both the state and the action spaces are high-dimensional spaces of variable size. In order to reduce the complexity of the decision-making problem, we have chosen to reduce these spaces so as to work with a small action space of constant size and a compact state space. In the following, we describe the procedure that was carried out for the reduction of the state and action spaces. 
	
	\subsection{Action space reduction: High-level actions} \label{sec: HLA}
	
	In this section, we elaborate on the design of a small and discrete set of actions that is an approximation of the original action space. Based on Assumptions (\ref{asu: OtherAgents}), (\ref{asu: ExogenousInfo}), (\ref{asu: StorageControl}), (\ref{asu: Aggressor}), (\ref{asu: Zero_Imbalances}) and (\ref{asu: DaysDecoupling}), a new action space $A'_{i}$ is proposed, which is defined as $A'_{i}=\{``Trade",``Idle" \}$. The new action space is composed of two high-level actions $a'_{i,t} \in A'_{i}$. These high-level actions are transformed to an original action through mapping $p: A'_{i} \rightarrow A_{i}^{red}$, from space $A'_{i}$ to the reduced action space $A_{i}^{red}$. The high-level actions are defined as following:
	
	\subsubsection{``Trade''}
	At each time-step $t$, agent $i$ selects orders from the order book with the objective of maximizing the instantaneous reward under the constraint that the storage device can remain balanced for every delivery period, even if no further interaction with the CID market occurs. As a reminder, this constraint was imposed by Assumption (\ref{asu: Zero_Imbalances}). 
	
	Under this assumption, the instantaneous reward signal $r_{i,t}$, presented in equation (\ref{eqn: reward}), consists only of the trading revenues obtained from the matching process of orders at time-step $t$. We will further assume that mapping $u:\mathbb{R}^{+} \times \{``Sell",``Buy"\} \rightarrow \mathbb{R}$ that adjusts the sign of the volume $v^{OB}$ of each order according to their side $y^{OB}$. Orders posted for buying energy will be associated with positive volume and orders posted for selling energy with negative volume, or equivalently:
	\begin{gather}
	u(v^{OB}, y^{OB}) = \begin{cases}
	v^{OB},& \text{if } y^{OB} = ``Buy",\\
	-v^{OB},& \text{if } y^{OB} = ``Sell".\end{cases} \label{eqn: volume}
	\end{gather}
	
	Consequently, the reward function $\rho$ defined in Section \ref{sec: Trading_rewards} is adapted according to the proposed modifications. The new reward function $\rho$, where $\rho : S^{OB} \times \bar{A}_{i}^{red} \to \mathbb{R}$, is a stationary function of the orders observed at each time-step $t$ and the agent's response to the observed orders. An analytical expression for the instantaneous reward collected is given by:
	\begin{gather}
	r_{i,t} =\rho\left(s^{OB}_{t},\bar{a}_{i,t} \right) = \sum_{j= 1}^{N_t} a^{j}_{i,t} \cdot u(v_{j}^{OB}, y_{j}^{OB}) \cdot p_{j}^{OB}. \label{eqn: Rewards}
	\end{gather} 
	
	The High-level action ``Trade" amounts to solving the bid acceptance optimisation problem presented in Algorithm~\ref{algo: Trade}. The objective function of the problem, formulated in equation (\ref{eqn: Obj}), consists of the revenues arising from trading. It is important to note that the operational constraints guarantee that no order will be accepted if it causes any imbalance. We denote as $N_{\tau} \subset \mathbb{N}$ the set of unique indices of the available orders that correspond to delivery time-step $\tau$ and $N_t = \bigcup_{\tau \in \bar{T}} N_{\tau}$. In equation (\ref{eqn: Balance}), the energy purchased and sold ($\sum_{j \in N_{\tau}} a_{i,t}^{j} u(v_{j}^{OB})$), the past net energy trades ($P_{i,t}^{mar}(\tau)$) and the energy discharged by the storage ($G_{i,t}(\tau)$) must match the energy charged by the storage ($C_{i,t}(\tau)$) for every delivery time-step $\tau$. The energy balance of the storage device, presented in equation (\ref{eqn: SoC_updates}), is responsible for the time-coupling and the arbitrage between two products $x$ (delivery time-steps $\tau$). The technical limits of the storage level and the charging and discharging process are described in equations (\ref{eqn: SoCLims}) to (\ref{eqn: DisLims}). The binary variables $k_{i,t} = (k_{i,t}(\tau), \forall \tau \in \bar{T} )$ restrict the operation of the unit for each delivery period in only one mode, either charging or discharging.
	
	The optimal solution to this problem yields the vector of fractions: $$\bar{a}_{i,t} = \left( a_{i,t}^{j}, \forall j \in N_t\right)\in \bar{A}_{i}^{red}$$ that are used in equation (\ref{eqn: ReducedAction}) to construct the action $a_{i,t}\in A_{i}^{red}$. The optimal solution also defines at each time-step $t$ the adjustments in the level of the production (discharge) $\Delta G_{i,t} = (\Delta G_{i,t}(\tau), \forall \tau \in \bar{T}(t))$ and the consumption (charge) $\Delta C_{i,t}= (\Delta C_{i,t}(\tau), \forall \tau \in \bar{T}(t))$. The evolution of the state of charge $SoC_{i,t+1}= (SoC_{i,t+1}(\tau), \forall \tau \in \bar{T}(t))$ of the unit as well as the production $G_{i,t+1} = ( G_{i,t+1}(\tau), \forall \tau \in \bar{T}(t))$ and consumption $C_{i,t+1} = ( C_{i,t+1}(\tau), \forall \tau \in \bar{T}(t))$ levels are computed for each delivery period.
	
	\begin{algorithm}
		\KwIn{$t$, $s_{t}^{OB}$, $P_{i,t}^{mar}$, $\underline{SoC_{i}}$, $\overline{SoC_{i}}$, $\underline{C_{i}}$, $\overline{C_{i}}$, $\underline{G_{i}}$, $\overline{G_{i}}$, $SoC_{i}^{init}$, $SoC_{i}^{term}$,$\tau_{init}$, $\tau_{term}$,$G_{i,t}$, $C_{i,t}$}
		\KwOut{$\bar{a}_{i,t}$, $SoC_{i,t+1}$, $G_{i,t+1}$, $C_{i,t+1}$, $\Delta G_{i,t}$, $\Delta C_{i,t}$, $k_{i,t+1}$, $r_{i,t}$}
		Solve:
		\begin{align}
		\hspace{-32pt}\underset{\tiny{\begin{array}{l} \bar{a}_{i,t},\mbox{ } SoC_{i,t+1} \\ G_{i,t+1},\mbox{ } C_{i,t+1}\\ \Delta G_{i,t},\mbox{ } \Delta C_{i,t}\\k_{i,t+1},\mbox{ } r_{i,t} \end{array}}} \max& \hspace{-5pt} \sum_{j \in N_t} a_{i,t}^{j} \cdot u(v_{j}^{OB}, y_{j}^{OB}) \cdot p_{j}^{OB}& \label{eqn: Obj}\\
		\hspace{-32pt}\text{s.t.} & \sum_{j \in N_{\tau}} a_{i,t}^{j} u(v_{j}^{OB}, y_{j}^{OB}) + P_{i,t}^{mar}(\tau) + \nonumber\\
		\hspace{-32pt}&\hspace{5pt} C_{i,t+1}(\tau) = G_{i,t+1}(\tau),&\hspace{-6pt}\forall \tau \in \bar{T}(t) \label{eqn: Balance}\\
		\hspace{-32pt}&SoC_{i,t+1}(\tau+ \Delta \tau) = SoC_{i,t+1}(\tau) +& \nonumber\\
		\hspace{-32pt}&\hspace{5pt} \Delta \tau \cdot \left( \eta \cdot C_{i,t+1}(\tau) - \frac{G_{i,t+1}(\tau)}{\eta}\right),&\hspace{-6pt}\forall \tau \in \bar{T}(t) \label{eqn: SoC_updates} \\
		\hspace{-32pt}	& \socmin{i} \leq SoC_{i,t+1}(\tau) \leq \socmax{i},&\hspace{-6pt}\forall \tau \in \bar{T}(t) \label{eqn: SoCLims}\\
		\hspace{-32pt}	& SoC_{i}^{init} = SoC_{i,t+1}(\tau_{init}),& \label{eqn: SoCinit}\\\hspace{-6pt}
		\hspace{-32pt}	& SoC_{i}^{term} = SoC_{i,t+1}(\tau_{term}),& \label{eqn: SoCterm}\\\hspace{-6pt}
		\hspace{-32pt}	& \underline{C_{i}} \leq C_{i,t+1}(\tau) \leq k_{i,t+1}(\tau)\cdot \overline{C_{i}},&\hspace{-6pt}\forall \tau \in \bar{T}(t) \label{eqn: PchLims}\\
		\hspace{-32pt}	& \underline{G_{i}} \leq G_{i,t+1}(\tau) \leq \left( 1-k_{i,t+1}(\tau) \right) \overline{G_{i}},&\hspace{-6pt}\forall \tau \in \bar{T}(t) \label{eqn: DisLims}\\
		\hspace{-32pt}	& G_{i,t+1}(\tau) = G_{i,t}(\tau) + \Delta G_{i,t}(\tau),&\hspace{-6pt}\forall \tau \in \bar{T}(t) \label{eqn: DisUpdates}\\
		\hspace{-32pt}	& C_{i,t+1}(\tau) = C_{i,t}(\tau) + \Delta C_{i,t}(\tau),&\hspace{-6pt}\forall \tau \in \bar{T}(t) \label{eqn: ChUpdates}\\
		\hspace{-32pt}	& k_{i,t+1}(\tau) \in \left\lbrace 0,1 \right\rbrace,&\hspace{-6pt} \forall \tau \in \bar{T}(t)\\
		\hspace{-32pt}	& a_{i,t}^{j} \in \left[ 0 , 1 \right],&\hspace{-6pt} \forall j \in N_t \label{eqn: Actions_space}
		\end{align} 
		\caption{``Trade"}
		\label{algo: Trade}
	\end{algorithm}
	
	\subsubsection{``Idle''}
	No transactions are executed, and no adjustment is made to the previously scheduled quantities. Under this action, the vector of fractions $\bar{a}_{i,t}$ is a zero vector. The discharge and charge as well as the state of charge of the storage device remain unchanged ($\Delta G_{i,t} \equiv 0$ and $\Delta C_{i,t} \equiv 0$) and we have:
	\begin{gather}
	G_{i,t+1}(\tau) = G_{i,t}(\tau),\forall \tau \in \bar{T}(t), \\
	C_{i,t+1}(\tau) = C_{i,t}(\tau),\forall \tau \in \bar{T}(t),\\
	SoC_{i,t+1}(\tau) = SoC_{i,t}(\tau),\forall \tau \in \bar{T}(t).\label{eqn: IdleSchedule}
	\end{gather}
	
	With such a reduction of the action-space, the agent can choose at every time-step $t$ between the two described high-level actions ($a'_{i,t} \in A'_{i} = \{ ``Trade",``Idle" \}$). Note that when the agent learns to idle, given a current situation, it does not necessarily mean, that if it had chosen to $``Trade"$ instead, he would not make a positive immediate reward. Indeed, the agent would choose $``Idle"$ if it believes that there may be a better market state emerging, i.e. the agent would learn to wait for the ideal opportunity of orders appearing in the order book at subsequent time-steps. We compare this approach to an alternative, which we refer to as the ``rolling intrinsic'' policy. According to this policy, at every time-step $t$ of the trading horizon the agent selects the combination of orders that optimises its operation and profits, based on the current information assuming that the storage device must remain balanced for every delivery period as presented in \cite{lohndorf2015optimal}. The ``rolling intrinsic'' policy is, thus, equivalent to sequentially selecting the action $``Trade"$ (Algorithm \ref{algo: Trade}), as defined in this framework. The algorithm proposed later in this paper exploits the experience that the agent can gain through (artificial) interaction with its environment, in order to learn the value of trading or idling at every different state that agent may encounter.
	
	\subsection{State space reduction}\label{sec: SSR}
	
	In this section, we propose a more compact and low-dimensional representation of the state space $H_{i}$. The state $h_{i,t}$, as explained in Section \ref{sec: Limitations}, contains the entire history of all the relevant information available for the decision-making up to time $t$. We consider each one of the components of the state $h_{i,t}$, namely the entire history of actions, order book states and private information, and we provide a simplified form.
	
	First, the vector containing the entire history of actions is reduced to a vector of binary variables after the modifications introduced in Section \ref{sec: HLA}.
	
	Second, the vector containing the history of order book states is reduced into a vector of engineered features. We start from the order book state $s^{OB}_{t} = ((x^{OB}_j,y'^{OB}_j,v^{OB}_j,p^{OB}_j,e^{OB}_j), \forall j \in N_{t} \subseteq \mathbb{N}) \in S^{OB} $ that is defined in Sections \ref{sec: CIDDesign} and \ref{sec: CIDEnv} as a high-dimensional continuous vector used to describe the state of the CID market. Owing to the variable (non-constant) and large amount of orders $\mid \hspace{-1.5pt} N_{t} \hspace{-1.5pt} \mid$, the space $S^{OB}$ has a non-constant size with high-dimensionality. 
	
\begin{figure}
		\centering
		\subfloat[\label{fig: Seperate_products_Liquid}]{\includegraphics[width=0.5\textwidth]{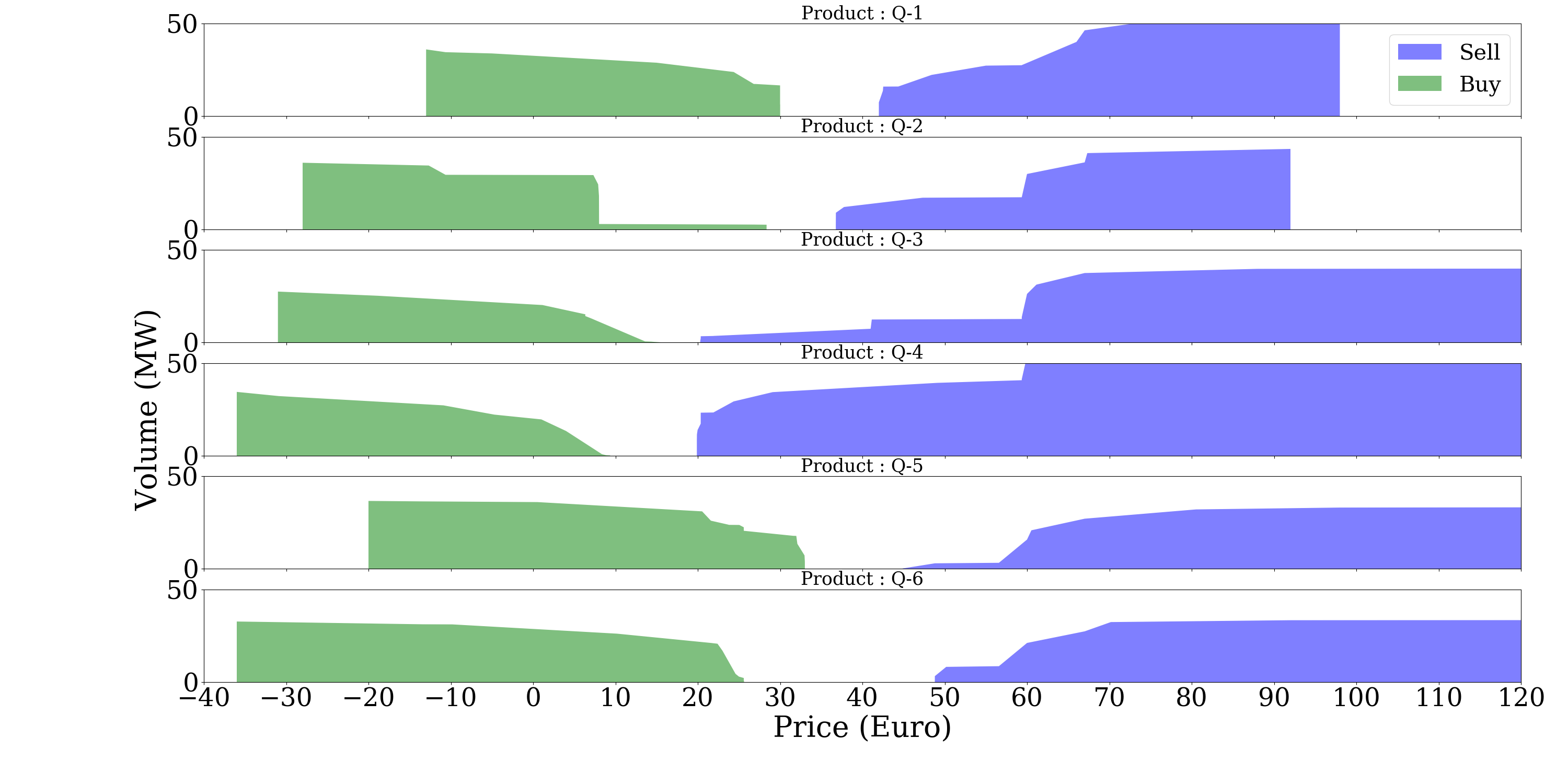}}\\
		\subfloat[\label{fig: Aggregate_products_Liquid}]{\includegraphics[width=1\linewidth]{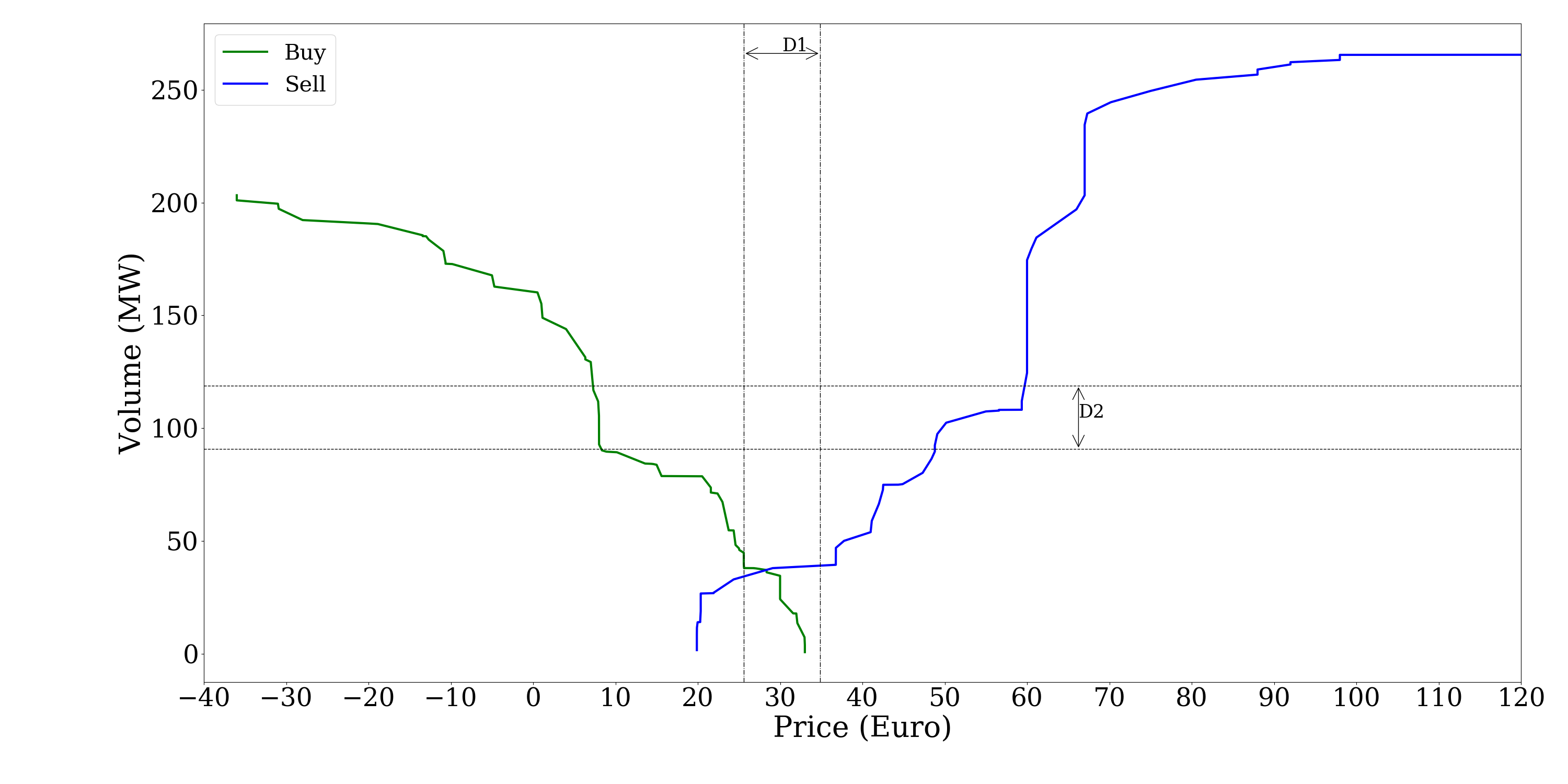}}
		\caption{(a) Market depth per product (for products $Q_1$ to $Q_6$) at a time-step $t$ with arbitrage potential. (b) The corresponding aggregated curves for a profitable order book.}
	\end{figure}

	In order to overcome this issue, we proceed as following. First, we consider the market depth curves for each product $x$. The market depth of each side (``Sell'' or ``Buy'') at a time-step $t$, is defined as the total volume available in the order book per price level for product $x$. The market depth for the ``Sell'' (``Buy'') side is computed by stacking the existing orders in ascending (descending) price order and accumulating the available volume. The market depth for each of the quarter-hourly products $Q_1$ to $Q_6$ at time instant $t$ is illustrated in Figure \ref{fig: Seperate_products_Liquid} using data from the German CID market. The market depth curves serve as a visualization of the order book that provides information about the liquidity of the market. Moreover, it provides information about the maximum (minimum) price that a trading agent will have to pay in order to buy (sell) a certain volume of energy. If we assume a fixed-price discretisation, certain upper and lower bounds on the prices and interpolation of the data in this price range, the market depth curves of each product $x$ can be approximated by a finite and constant set of values.
	
	Even though this set of values has a constant size, it can still be extremely large. Its dimension is not a function of the number of existing orders any more, but it depends on the resolution of the price discretisation, the price range considered, and the total number of products in the market. Instead of an individual market depth curve for each product $x$, we consider a market depth curve for all the available products, i.e. existing orders in ascending (descending) price order and accumulating the available volumes for all the products. In this way we can construct the aggregated market depth curve, presented in Figure \ref{fig: Aggregate_products_Liquid}. The aggregated market depth curve illustrates the total available volume (``Sell'' or ``Buy'') per price level for all products.
	
	The motivation for considering the aggregated curves comes from the very nature of a storage device. The main profit-generating mechanism of a storage device is the arbitrage between two delivery periods. Its functionality involves the purchasing (charging) of electricity during periods of low prices and the selling (discharging) during periods of high prices. 
	
\begin{figure}
		\centering
		\subfloat[\label{fig: Seperate_products_NonLiquid}]{\includegraphics[width=1\linewidth]{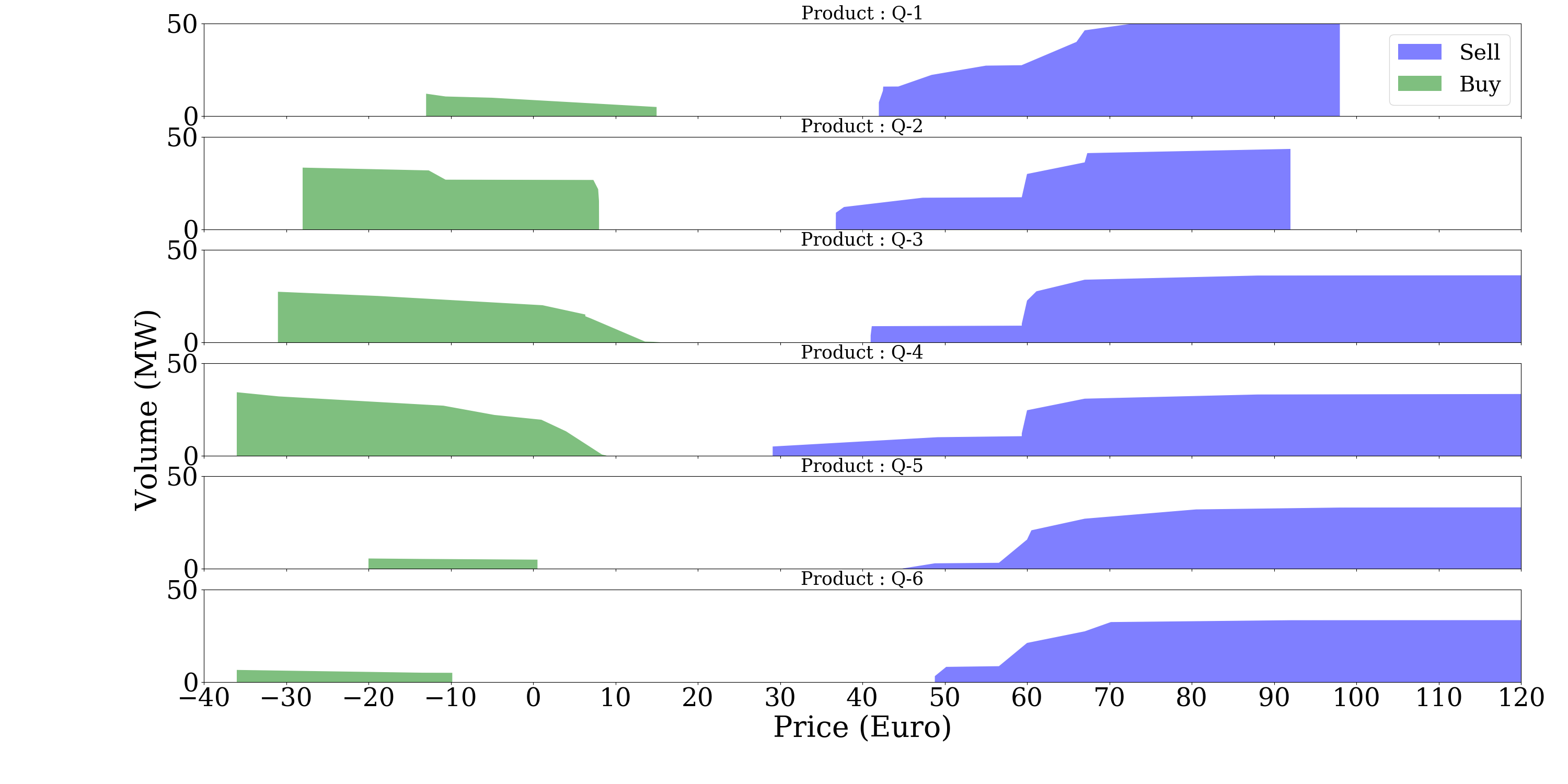}}\\
		\subfloat[\label{fig: Aggregate_products_NonLiquid}]{\includegraphics[width=1\linewidth]{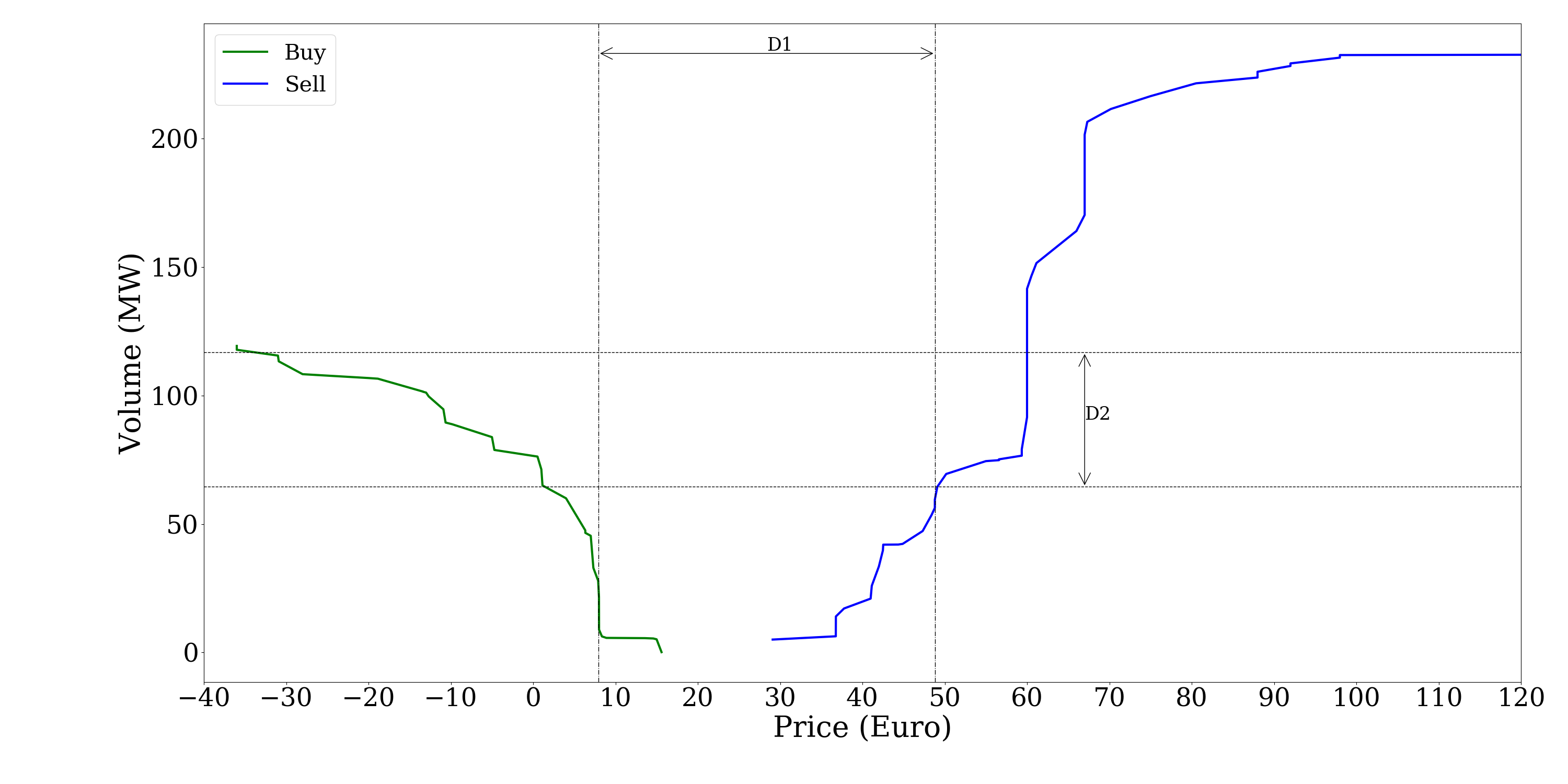}}
		\caption{(a) Market depth per product (for products $Q_1$ to $Q_6$) at a time-step $t$ with no arbitrage potential. (b) The corresponding aggregated curves for a non profitable order book.}
	\end{figure}

	For instance, in Figure \ref{fig: Seperate_products_Liquid}, a storage device would buy volume for product $Q_4$ and sell volume back for product $Q_5$. The intersection of the ``Sell'' and ``Buy'' curves in Figure \ref{fig: Aggregate_products_Liquid} defines the maximum volume that can be arbitraged by the storage device if no operational constraints were considered and serves as an upper bound for the profits at each step $t$. Alternatively, the market depth for the same products $Q_1$ to $Q_6$ at a different time-step of the trading horizon is presented in Figure \ref{fig: Seperate_products_NonLiquid}. As illustrated in Figure \ref{fig: Aggregate_products_NonLiquid}, there is no arbitrage opportunity between the products, hence the aggregated curves do not intersect. Thus, we assume, that the aggregated curves provide a sufficient representation of the order book.
	
	At this point, considering a fixed-price discretisation and a fixed price range would yield a constant set of values able to describe the aggregated curves. However, in order to further decrease the size of the set of values with sufficient price discretisation, we motivate the use of a set of distance measures between the two aggregated curves that succeed in capturing the arbitrage potential at each trading time-step $t$ as state variables, as presented in Figures \ref{fig: Aggregate_products_Liquid} and \ref{fig: Aggregate_products_NonLiquid}. 
	
	For instance, we define as $D1$ the signed distance between the 75th percentile of ``Buy'' price and the 25th percentile of ``Sell'' price and as $D2$ the absolute distance between the mean value of ``Buy'' and ``Sell'' volumes. Other measures used are the signed price difference and absolute volume difference between percentiles (25\%, 50\%, 75\%) and the bid-ask spread. A detailed list of the distance measures is provided in Table \ref{table: Order_Book_reduction}.
	
	\begin{table*}[h]
		\centering
		\caption{Order book features used for the state reduction.}
		\renewcommand{\arraystretch}{1.5}
		
		\label{table: Order_Book_reduction}
		\begin{tabular}{llll}
			\textbf{Symbol} 	& \textbf{Definition} 					& \textbf{Description} &\\ \cmidrule[1pt]{1-4} 
			$D1$ 				& $p_{max}^{Buy} - p_{min}^{Sell} $		& Signed diff. between the maximum ``Buy'' price and the minimum ``Sell'' price										&\\ 
			$D2$ 				& $p_{mean}^{Buy} - p_{mean}^{Sell}$ 	& Signed diff. between the mean ``Buy'' price and the mean ``Sell'' price	 											&\\
			$D3$ 				& $p_{25\%}^{Buy} - p_{75\%}^{Sell}$ 	& Signed diff. between the 25th percentile ``Buy'' price and the 75th percentile ``Sell'' price	 					&\\
			$D4$ 				& $p_{50\%}^{Buy} - p_{50\%}^{Sell}$ 	& Signed diff. between the 50th percentile ``Buy'' price and the 50th percentile ``Sell'' price	 					&\\
			$D5$ 				& $p_{75\%}^{Buy} - p_{25\%}^{Sell}$ 	& Signed diff. between the 75th percentile ``Buy'' price and the 25th percentile ``Sell'' price	 					&\\
			$D6$ 				& $|v_{min}^{Buy} - v_{min}^{Sell}|$	& Abs. diff. between the minimum ``Buy'' cum. volume and the maximum ``Sell'' cum. volume 						&\\ 
			$D7$ 				& $|v_{mean}^{Buy} - v_{mean}^{Sell}|$ 	& Abs. diff. between the mean ``Buy'' cum. volume and the mean ``Sell'' cum. volume 								&\\
			$D8$ 				& $|v_{25\%}^{Buy} - v_{25\%}^{Sell}|$ 	& Abs. diff. between the 25th percentile ``Buy'' cum. volume and the 25th percentile ``Sell'' cum. volume 		&\\
			$D9$ 				& $|v_{50\%}^{Buy} - v_{50\%}^{Sell}|$ & Abs. diff. between the 50th percentile ``Buy'' cum. volume and the 50th percentile ``Sell'' cum. volume 		&\\
			$D10$ 				& $|v_{75\%}^{Buy} - v_{75\%}^{Sell}|$ & Abs. diff. between the 75th percentile ``Buy'' cum. volume and the 75th percentile ``Sell'' cum. volume 		&
		\end{tabular}
	\end{table*}
	
	The new, continuous, low-dimensional observation of the order book $s_t'^{OB} \in S'^{OB} = \{D1,..,D10 \}$ is used to represent the state of the order book and, in particular, its profit potential. It is important to note that in contrast to $s_t^{OB} \in S^{OB}$, the new order book observation $s_t'^{OB}\in S'^{OB}$ does not depend on the number of orders in the order book and therefore has a constant size, i.e. the cardinality of $S'^{OB}$ is constant over time.
	
	Finally, the history of the private information of agent $i$, that is not publicly available, is a vector that contains the high-dimensional continuous variables $s^{private}_{i,t}$ related to the operation of the storage device. As described in Section \ref{sec: Limitations}, $s^{private}_{i,t}$ is defined as:
	\begin{align}
	s^{private}_{i,t} = &((P_{i,t}^{mar}(\tau),\Delta_{i,t}(\tau),\nonumber\\
	&G_{i,t}(\tau),C_{i,t}(\tau),SoC_{i,t}(\tau),\forall \tau \in \bar{T}),\nonumber\\
	&w^{exog}_{i,t}).\nonumber
	\end{align} 
	
	According to Assumption (\ref{asu: Zero_Imbalances}), the trading agent cannot perform any transaction if it results in imbalances. Therefore, it is not relevant to consider the vector $\Delta_{i,t}$ since it will always be zero according to the way the high-level actions are defined in Section \ref{sec: HLA}. Additionally, Assumption (\ref{asu: StorageControl}) regarding the default strategy for storage control in combination with Assumption (\ref{asu: Zero_Imbalances}) yields a direct correlation between vectors $P_{i,t}^{mar}$ and $G_{i,t}$, $C_{i,t}$, $SoC_{i,t}$. Thus, it is considered that $P_{i,t}^{mar}$ contains all the required information and thus vectors $G_{i,t}$, $C_{i,t}$ and $SoC_{i,t}$ can be dropped.
	
	Following the previous analysis we can define the low-dimensional pseudo-state $z_{i,t}= ( s'_{i,0},a'_{i,0},r_{i,0},...,a'_{i,t-1},r_{i,t-1},s'_{i,t} ) \in Z_{i}$, where $s'_{i,t} = (s'^{OB}_t,P_{i,t}^{mar},w^{exog}_{i,t}) \in S'_{i}$. This pseudo-state can be seen as the result of applying an encoder $e: H_{i} \rightarrow Z_{i}$ which maps a true state $h_{i,t}$ to pseudo-state $z_{i,t}$.


	\begin{asu}[\emph{Pseudo-state}]\label{asu: PseudoState}
		The pseudo-state $z_{i,t} \in Z_{i}$ contains all the relevant information for the optimisation of the CID market trading of an asset-optimizing agent.
	\end{asu}

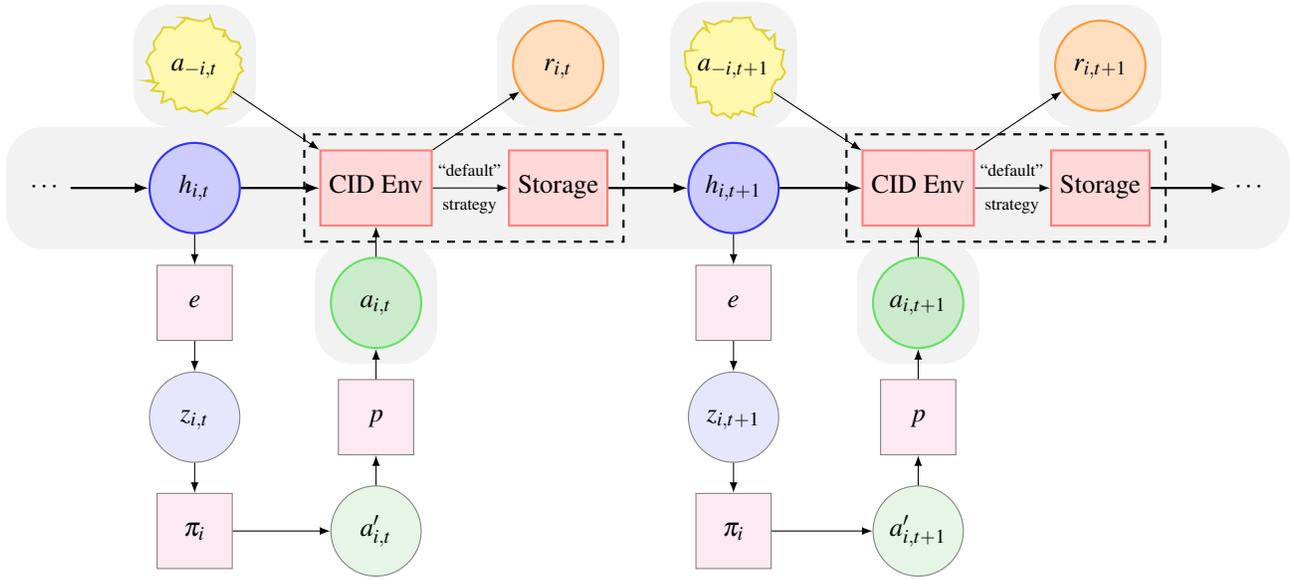
\begin{figure*}
		\centering
		\tikzstyle{state}=[circle,
		thick,
		minimum size=1.2cm,
		draw=blue!80,
		fill=blue!20]
		
		\tikzstyle{measurement}=[circle,
		thin,
		minimum size=1.2cm,
		draw=black!50,
		fill=blue!10]
		
		\tikzstyle{reward}=[circle,
		thick,
		minimum size=1.2cm,
		draw=orange!80,
		fill=orange!25
		]
		
		\tikzstyle{env}=[rectangle,
		thick,
		minimum size=1cm,
		draw=red!50,
		fill=red!15]
		
		\tikzstyle{map}=[rectangle,
		minimum size=1cm,
		draw=black!50,
		fill=magenta!10]
		
		\tikzstyle{noise}=[circle,
		thick,
		minimum size=1.2cm,
		draw=yellow!85!black,
		fill=yellow!40,
		decorate,
		decoration={random steps,
			segment length=2pt,
			amplitude=2pt}]
		
		\tikzstyle{action}=[circle,
		thick,
		minimum size=1.2cm,
		draw=green!80!black!60,
		fill=green!60!black!20]
		
		\tikzstyle{action_1}=[circle,
		thin,
		minimum size=1.2cm,
		draw=black!50,
		fill=green!60!black!10]
		
		\tikzstyle{background_1}=[rectangle,
		fill=gray!10,
		inner sep=0.2cm,
		rounded corners=5mm]
		
		\tikzstyle{background_2}=[rectangle,
		thick, dashed,
		draw=black!100,
		inner sep=0.2cm,
		rounded corners=0mm]
		
		\begin{tikzpicture}[>=latex,text height=1.ex,text depth=0.25ex]
		
		\matrix[row sep=0.3cm,column sep=1.0cm] {
			
			
			& \node (w_k) [noise] {${a_{-i,t}}$}; & 	 	 &	 	\node (r_k) [reward] {$r_{i,t}$}; 		& \node (w_k+1) [noise] {${a_{-i,t+1}}$}; & 	& \node (r_k+1) [reward] {$r_{i,t+1}$}; 		&
			\\
			\node (A_k-2)  {$\cdots$};  & \node (x_k) [state] {${h_{i,t}}$}; & \node (A_k) [env] {CID Env}; & \node (B_k) [env] {Storage}; & \node (x_k+1) [state] {${h_{i,t+1}}$}; & \node (A_k+1) [env] {CID Env}; & \node (B_k+1) [env] {Storage}; & \node (x_k+2) {$\cdots$};
			\\
			
			&	\node (H_k) [map] {${e}$}; & 	\node (a_k) [action] {${a_{i,t}}$};		 &					 &	\node (H_k+1) [map] {${e}$}; &	 	\node (a_k+1) [action] {${a_{i,t+1}}$};		 &	&
			\\
			
			&	\node (z_k) [measurement] {${z_{i,t}}$}; & 	\node (p_k) [map] {${p}$};			&	&\node (z_k+1) [measurement] {${z_{i,t+1}}$}; & \node (p_k+1) [map] {${p}$};			& &
			\\
			&	\node (pi_k) [map] {${\pi_{i}}$}; & 	\node (a'_k) [action_1] {${a'_{i,t}}$};			& & \node (pi_k+1) [map] {${\pi_{i}}$};	 & \node (a'_k+1) [action_1] {${a'_{i,t+1}}$}; & &
			\\
		};
		
		\path[->]
		(A_k-2) edge[thick] (x_k)	
		(x_k) edge[thick] (A_k)		
		(B_k) edge[thick] (x_k+1)	
		(x_k+1) edge[thick] (A_k+1)
		
		(A_k) edge node[below] {\hspace{-3pt}{\fontsize{6.5pt}{48} \selectfont \textcolor{black}{strategy}}} node[above] {\hspace{-3pt}{\fontsize{6.5pt}{48} \selectfont \textcolor{black}{``default"}}} (B_k)
		
		(x_k) edge (H_k)
		(H_k) edge (z_k)
		(x_k+1) edge (H_k+1)
		(H_k+1) edge (z_k+1)
		
		(a'_k) edge (p_k) 
		(p_k) edge (a_k)
		(a_k) edge (A_k)

		(w_k) edge (A_k)
		
		(z_k) edge (pi_k)
		(pi_k) edge (a'_k)
		(A_k) edge (r_k)
		(w_k+1) edge (A_k+1)
		(z_k+1) edge (pi_k+1)
		(pi_k+1) edge (a'_k+1)
		
		(a'_k+1) edge (p_k+1) 
		(p_k+1) edge (a_k+1)
		(a_k+1) edge (A_k+1)
		
		(A_k+1) edge node[below] {\hspace{-3pt}{\fontsize{6.5pt}{48} \selectfont \textcolor{black}{strategy}}} node[above] {\hspace{-3pt}{\fontsize{6.5pt}{48} \selectfont \textcolor{black}{``default"}}} (B_k+1)
		(B_k+1) edge[thick] (x_k+2)
		
		(A_k+1) edge (r_k+1)
		;

		\begin{pgfonlayer}{background}
		\node [background_1,
		fit= (A_k-2) (x_k) (A_k) (x_k+1) (A_k+1) (x_k+2) ]{};
		\node [background_1,
		fit=(a_k)] {};
		\node [background_1,
		fit=(r_k)] {};
		\node [background_1,
		fit=(w_k)] {};
		
		\node [background_1,
		fit=(a_k+1)] {};
		\node [background_1,
		fit=(r_k+1)] {};
		\node [background_1,
		fit=(w_k+1)] {};
		
		\node [background_2,
		fit=(A_k) (B_k)] {};
		
		\node [background_2,
		fit=(A_k+1) (B_k+1)] {};
		\end{pgfonlayer}
		\end{tikzpicture}
		
		\caption{Schematic of the decision process. The original MDP is highlighted in a gray background. The state of the original MDP $h_{i,t}$ is encoded in pseudo-state $z_{i,t}$. Based on $z_{i,t}$, agent $i$ takes an high-level action $a'_{i,t}$, according to its policy $\pi_{i}$. This action $a'_{i,t}$ is mapped to an original action $a_{i,t}$ and submitted to the CID market. The CID market makes a transition based on the action of agent $i$ and the actions of the other agents $a_{-i,t}$. After this transition, the market position of agent $i$ is defined and the control actions for storage device are derived according to the ``default" strategy. Each transition yields a reward $r_{i,t}$ and a new state $h_{i,t}$.}\label{fig: MDP}
	\end{figure*}

	Under Assumption (\ref{asu: PseudoState}), using the pseudo-state $z_{i,t}$ instead of the true state $h_{i,t}$ is equivalent and does not lead to a sub-optimal policy. The resulting decision process after the state and action spaces reductions is illustrated in Figure \ref{fig: MDP}.

	

	\begin{figure*}
		\centering
		\tikzstyle{hidden}=[circle,
		thin,
		minimum size=0.9cm,
		draw=orange!80,
		fill=orange!25]
		
		\tikzstyle{measurement}=[circle,
		thin,
		minimum size=1.2cm,
		draw=black!50,
		fill=blue!10]
		
		\tikzstyle{reward}=[circle,
		thin,
		minimum size=1.2cm,
		draw=orange!25,
		fill=orange!25
		]
		
		\tikzstyle{state}=[rectangle,
		thin,
		minimum size=0.7cm,
		draw=black!30,
		fill=green!60!black!10]
		
		\tikzstyle{map}=[rectangle,
		minimum size=1cm,
		draw=black!30,
		fill=green!60!black!20]
		
		\tikzstyle{noise}=[circle,
		thick,
		minimum size=1.2cm,
		draw=yellow!85!black,
		fill=yellow!40,
		decorate,
		decoration={random steps,
			segment length=2pt,
			amplitude=2pt}]
		
		\tikzstyle{action}=[circle,
		thick,
		minimum size=1.2cm,
		draw=green!80!black!60,
		fill=green!60!black!20]
		
		\tikzstyle{action_1}=[circle,
		thin,
		minimum size=1.2cm,
		draw=black!50,
		fill=green!60!black!10]
		
		\tikzstyle{background_1}=[rectangle,
		fill=gray!10,
		inner sep=0.2cm,
		rounded corners=5mm]
		
		\tikzstyle{background_2}=[rectangle,
		fill=yellow!15,
		inner sep=0.2cm,
		rounded corners=2mm]
		
		\tikzstyle{background_3}=[rectangle,
		thick, dashed,
		draw=black!100,
		inner sep=0.2cm,
		rounded corners=0mm]
		
		\begin{tikzpicture}[>=latex,text height=1.ex,text depth=0.25ex]
		
		\matrix[row sep=0.3cm,column sep=0.5cm] {
			
			\node (O)  {Output};			&&	       							 					&   													&  	\node (Q_0) [measurement] {${Q_{Trade}}$};		&		  										 &  \node (Q_1) [measurement] {${Q_{Idle}}$};			& 	\\
			&	&	       							 					& \node (I) {\fontsize{9pt}{48} \selectfont \textcolor{black}{FC5 shape: (batch size, 36, 2)}} ;&  	\node (F_11) [state] {${F}$};					&	\node (F_12) [state] {${F}$}; 				& \node (Dots_3)  {$\cdots$};  & \node (F_13) [state] {${F}$};	\\
			
			\node (F)  {Fully Connected};	&&	       							 					&  													 &  	\node (Dots_1)  {$\vdots$};						&		\node (Dots_2)  {$\vdots$};				 & &	\node (Dots_3)  {$\vdots$};\\
			
			&	&	       							 					&  \node (I) {\fontsize{9pt}{48} \selectfont \textcolor{black}{FC2 shape: (batch size, 36, 36)}} ;	&  	\node (F_21) [state] {${F}$};			 		&	\node (F_22) [state] {${F}$}; 				& \node (Dots_4)  {$\cdots$};	& \node (F_23) [state] {${F}$};\\
			
			&	&	       							 					&\node (I) {\fontsize{9pt}{48} \selectfont \textcolor{black}{FC1 shape: (batch size, 128, 36)}} ;	&  	\node (F_31) [state] {${F}$};			 		&	\node (F_32) [state] {${F}$};				 & \node (Dots_5)  {$\cdots$};	& \node (F_33) [state] {${F}$};\\
			
			\node (H)  {Hidden state};	&		&	\node (h_0) [hidden] {${h_{i,0}}$};					 & 	\node (h_1) [hidden] {${h_{i,1}}$};				&			\node (Dots_6)  {$\cdots$}; 			 & \node (h_t) [hidden] {${h_{i,t}}$};	 &  & \\	
			\node (Lstm)  {LSTM}; &\node (h_10) [hidden] {${h_{i,-1}}$};			&	\node (L_0) [map] {LSTM};							 & 	\node (L_1) [map] {LSTM};						&			\node (Dots_7)  {$\cdots$};										&\node (L_t) [map] {LSTM}; && \\
			\node (I)  {Inputs};		&	&	\node (z_0) [measurement] {${\bar{s}_{i,t-\bar{h}}}$};			 & 	\node (z_1) [measurement] {${\bar{s}_{i,t-\bar{h}+1}}$};		&			\node (Dots_8)  {$\cdots$}; 			 & \node (z_t) [measurement] {$\bar{s}_{i,t}$};	 && \\
			&	&	 													 & 	\node (I) {\fontsize{9pt}{48} \selectfont \textcolor{black}{Input shape: (batch size, $\bar{h}$, 263)}} ;							&		 										 &  								 && \\
		};
		
		%
		%
		%
		%
		%
		%
		%
		%
		%
		
		\path[->]
		(z_0) edge (L_0)	
		(z_1) edge (L_1)
		(z_t) edge (L_t)
		
		(L_0) edge (h_0)	
		(L_1) edge (h_1)
		(L_t) edge node[right] {\hspace{10pt}{\fontsize{9pt}{48} \selectfont \textcolor{black}{Lstm output: (batch size, 128)}}}  (h_t)
		
		(h_10) edge (L_0)
		(L_0) edge (L_1)	
		(L_1) edge (Dots_7)
		(Dots_7) edge (L_t)
		
		(h_t) edge (F_31.south)	
		(h_t) edge (F_32.south)
		(h_t) edge  (F_33.south)	
		
		(F_31.north) edge[ultra thin] (F_21.south)	
		(F_31.north) edge[ultra thin] (F_22.south)
		(F_31.north) edge[ultra thin] (F_23.south)
		
		(F_32.north) edge[ultra thin] (F_21.south)	
		(F_32.north) edge[ultra thin] (F_22.south)
		(F_32.north) edge[ultra thin] (F_23.south)
		
		(F_33.north) edge[ultra thin] (F_21.south)	
		(F_33.north) edge[ultra thin] (F_22.south)
		(F_33.north) edge[ultra thin] (F_23.south)
		
		(F_11.north) edge[ultra thin] (Q_0.south)	
		(F_12.north) edge[ultra thin] (Q_0.south)
		(F_13.north) edge[ultra thin] (Q_0.south)
		
		(F_11.north) edge[ultra thin] (Q_1.south)	
		(F_12.north) edge[ultra thin] (Q_1.south)
		(F_13.north) edge[ultra thin] (Q_1.south)
		%
		
		;

		\begin{pgfonlayer}{background}
		\node [background_1,
		fit= (F_11) (F_12) (F_13) (F_21) (F_22) (F_23) (F_31) (F_32) (F_33) ]{};
		
		
		\end{pgfonlayer}
		\end{tikzpicture}
		
		\caption{Schematic of the neural network architecture.}\label{fig: NN}
	\end{figure*}
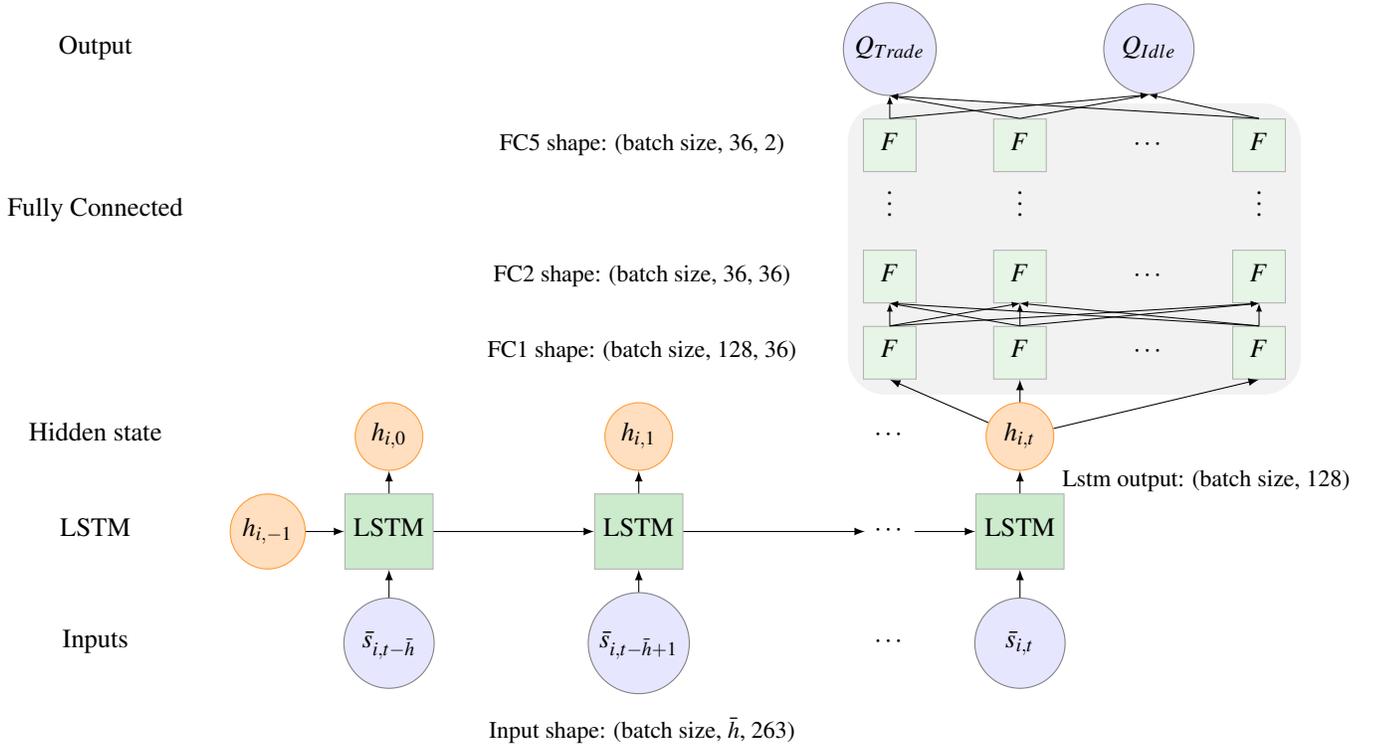

	\subsection{Generation of artificial trajectories} \label{sec: Trajectories_Generation}
	\begin{algorithm*}
		
		\KwIn{$L^{train}$, $E$, $ep$, $\epsilon$, $decay$}
		
		\KwOut{$\hat{Q}$, $F$}
		Initialize $\hat{Q}\equiv 0 $\;
		
		$M \gets E \cdot |L^{train}|$
		
		$m \gets 0$

		\Repeat{$m \ge M$} {
			\For{$iter_{j}\gets0$ \KwTo $ep$}{ 
				$d \gets rand(L^{train})$ \hfill $\vartriangleright$ Randomly pick a day $d$ from train set $L^{train}$
				
				$ \zeta_{m} \gets simulate(d, \epsilon-greedy(\hat{Q}))$ \hfill $\vartriangleright$ Generate trajectory $ \zeta_{m}$ by simulating day $d$ using $\epsilon$-greedy policy
				
				$F.add( \zeta_{m})$ \hfill $\vartriangleright$ Append trajectory from day $d$ to set $F$
				
				$\epsilon \gets anneal(\epsilon, decay, iter_{i})$ \hfill $\vartriangleright$ Anneal the value of $\epsilon$ based on $decay$ parameter
				
				$m \gets m + 1$
			}
			
			Update $\hat{Q}$ using set $F$ according to equations (\ref{eqn: LS}), (\ref{eqn: targets}) \hfill $\vartriangleright$ Fit new $\hat{Q}$ functions
			
		}
		\Return{$\hat{Q}, F$}\;
		\caption{Generation of artificial trajectories}
		\label{algo: FQI}
	\end{algorithm*}
	
	In this section, the generation of additional artificial trajectories for addressing exploration issues is discussed. Indeed, if we were to implement an agent that selects at every time-step among the ``Idle" and ``Trade" actions, we would collect a certain number of trajectories (one per day) over a certain period of interactions. The collected dataset could be used to train a policy using a batch mode RL algorithm, as described in Section \ref{sec: Solution}. Every time a new trajectory would arrive, it would be appended in the previous set of trajectories and the entire dataset could be used to improve the trading policy. 
	
	This approach requires a large number of days in order to acquire a sufficient amount of data. Additionally, as discussed in Section \ref{sec: Solution}, sufficient exploration of the state and action spaces is a key requirement for converging to a near-optimal policy. It is required that all parts of the state and action spaces are visited sufficiently often. As a result, the RL agent needs to explore unknown grounds in order to discover interesting policies (exploration). It should also apply these learned policies to get high rewards (exploitation). However, since the set of collected trajectories would come from a real agent, the visitation of many different states is expected to be limited. 
	
	In the RL context, exploration is performed when the agent selects a different action than the one that, according to its experience, will yield the highest rewards. In real life, it is unlikely for a trader to select such actions, and potentially bear negative revenues, for the sake of gaining more experience. This leads to limited exploration of the learning process and would result in a suboptimal policy.
	
	\begin{asu}[\emph{No impact on the behaviour of the rest of the agents}]\label{asu: Rest_behaviour}
		The actions of trading agent $i$ do not influence the future actions of the rest of the agents $-i$ in the CID market. In this way, agent $i$ is not capable of influencing the market.
	\end{asu}
	
	Assumption (\ref{asu: Rest_behaviour}) implies that each of the agents $-i$ entering in the market would post orders solely based on their individual needs. Furthermore, its actions are not considered as a reaction to the actions of the other market players. 
	
	Leveraging Assumption (\ref{asu: Rest_behaviour}) allows one to tackle the exploration issues discussed previously by generating several artificial trajectories using historical order book data. We denote by $E$ the number of episodes (times) each day from historical data is repeated and by $L^{train}$ the set of trading days used to train the agent. We can then obtain the total number of trajectories $M$ as $M =E \cdot |L^{train}|$.
	
	The generation of artificial trajectories is performed according to the process described in \cite{ernst2005tree}. This process interleaves the generation of trajectories with the computation of an approximate Q-function using the trajectories already generated. As shown in Algorithm \ref{algo: FQI}, for a number of episodes $ep$, we randomly select days from the training set which we simulate using an $\epsilon$-greedy policy. According to this policy, an action is chosen at random with probability $\epsilon$ and according to the available Q-functions with probability ($1-\epsilon$). The generated trajectories are added to the set of trajectories. The second step consists of updating the Q-function approximation using the set of collected trajectories. This process is terminated when the total number of episodes has reached the specified number $E$. 
	
	This process introduces parameters $L^{train}$, $E$, $ep$, $\epsilon$ and $decay$. The selection of these parameters impacts the training progress and the quality of the resulting policy. The set of days considered for training ($L^{train}$) is typically selected as a proportion (e.g. 70\%) of the total set of days available. The total number of episodes $E$ should be large enough so that convergence is achieved and is typically tuned based on the application. The frequency with which the trajectory generation and the updates are interleaved is controlled by parameter $ep$. A small number of $ep$ results in a large number of updates. Parameter $\epsilon$ is used to address the trade-off between exploration-exploitation during the training process. As the training evolves, this parameter is annealed based on some predefined parameter $decay$, in order to gradually reduce exploration and to favour exploration along the (near-)optimal trajectories. In practice, the size of the buffer $F$ cannot grow infinitely due to memory limitations, so typically a limit on the number of trajectories stored in the buffer is imposed. Once this limit is reached, the oldest trajectories are removed as new ones arrive. The buffer is a double-ended queue of fixed size.
	
	\subsection{Neural Network architecture}\label{sec: Network_architecture}

	As described in Section \ref{sec: SSR}, pseudo-state $z_{i,t}$ contains a sequence of variables whose length is proportional to $t$. This motivates the use of Recurrent Neural Networks (RNNs), that are known for being able to efficiently process variable-length sequences of inputs. In particular, we use Long Short-term Memory (LSTM) networks \cite{Goodfellow-et-al-2016}, a type of RNNs where a gating mechanism is introduced to regulate the flow of information to the memory state. 
	
	All the networks in this study have the architecture presented in Figure \ref{fig: NN}. It is composed of one LSTM layer with 128 neurons followed by five fully connected layers with 36 neurons where ``ReLU" was selected as the activation function. The structure of the network (number of layers and neurons) was selected after cross-validation. 
	
	Theoretically, the length of the sequence of features that is provided as input to the neural network can be as large as the total number of trading steps in the optimisation horizon. In practice though, there are limitations with respect to the memory that is required to store a tensor of this size. As we can observe in Figure \ref{fig: NN}, each sample in the batch contains a vector of size 263 for each time-step. Assuming a certain batch size, there is a certain limit to the number of steps that can be stored in the memory. Therefore, for practical reasons and due to hardware limitations, we assume a history length $\bar{h}$ defined as $z_{i,t} = (a'_{i,t-\bar{h}-1},r_{i,t-\bar{h}-1}, s'_{i,t-\bar{h}},a'_{i,t-\bar{h}},r_{i,t-\bar{h}},...,a'_{i,t-1},r_{i,t-1},s'_{i,t}) \in Z_{i}$. At each step $t$, the history length $\bar{h}$ takes the minimum value between the time-step $t$ and $\bar{h}_{max}$, ($\bar{h}=min(t, \bar{h}_{max})$). Additionally, we provide the variable $\bar{s}_t=(a'_{i,t-1},r_{i,t-1},s'_{i,t})$, as a fixed size input for each step $t$ of the LSTM. Consequently, the pseudo-state can be written as $z_{i,t} = ( \bar{s}_{t-\bar{h}}, ... , \bar{s}_t )$.
	
	\subsection{Asynchronous Distributed Fitted Q iteration}\label{sec: Async}
	\begin{figure*}
		\centering
		\tikzstyle{decision} = [diamond, draw, fill=blue!20, 
		text width=4.5em, text badly centered, node distance=3cm, inner sep=0pt]
		\tikzstyle{map}=[rectangle,
		minimum size=1cm,
		draw=black!30,
		fill=green!60!black!20]
		
		\tikzstyle{train} = [rectangle, draw=black!30,
		fill=green!60!black!20, 
		text width=7em, text centered, minimum height=7em]
		
		\tikzstyle{actor} = [rectangle, draw=red!50,
		fill=red!15, text width=7em, text centered, minimum height=7em]
		
		\tikzstyle{buffer} = [rectangle, draw, fill=blue!10, 
		text width=7em, text centered, minimum height=7em]

		\tikzstyle{line} = [draw, -latex']
		\tikzstyle{cloud} = [draw, ellipse,fill=red!20, node distance=3cm,
		minimum height=2em]
		
		\begin{tikzpicture}[node distance = 2cm, auto]
		\node [train] (train) {\underline{Learner} \\ Network};
		
		\node [actor, right of=train, below of=train, node distance=4.2cm] (actor1) {};
		\node [actor, right of=train, below of=train, node distance=4.3cm] (actor2) {};
		\node [actor, right of=train, below of=train, node distance=4.4cm] (actor3) {};
		\node [actor, right of=train, below of=train, node distance=4.5cm] (actor) {\underline{Actor} \\ Network \\ Environment \\ Local Buffer};
		\node [buffer, right of=actor, above of=actor, node distance=4.5cm] (buffer) {\underline{Global buffer} \\ Experiences};

		
		\path [line] (train) |- node[near end] {\hspace{-5pt}\vspace{0pt}{\fontsize{9pt}{48} \selectfont \textcolor{black}{Network parameters}}} (actor) ;
		
		\path [line] (actor) -| node[near start] {\hspace{0pt}{\fontsize{9pt}{48} \selectfont \textcolor{black}{Local buffer}}} (buffer);
		
		\path [line] (buffer) -- node[above] {\hspace{0pt}{\fontsize{9pt}{48} \selectfont \textcolor{black}{Experiences}}} (train);
		
		\end{tikzpicture}
		\caption{Schematic of the asynchronous distributed architecture. Each actor runs on a different thread and contains a copy of the environment, an individual $\epsilon$-greedy policy based on the latest version of the network parameters and a local buffer. The actors generate trajectories that are stored in their local buffers. When the local buffer of each actor is filled, it is appended to the global buffer and the agent collects the latest network parameters from the learner. A single learner runs on a separate thread and is continuously training using experiences from the global buffer.}\label{fig: Training}
	\end{figure*}
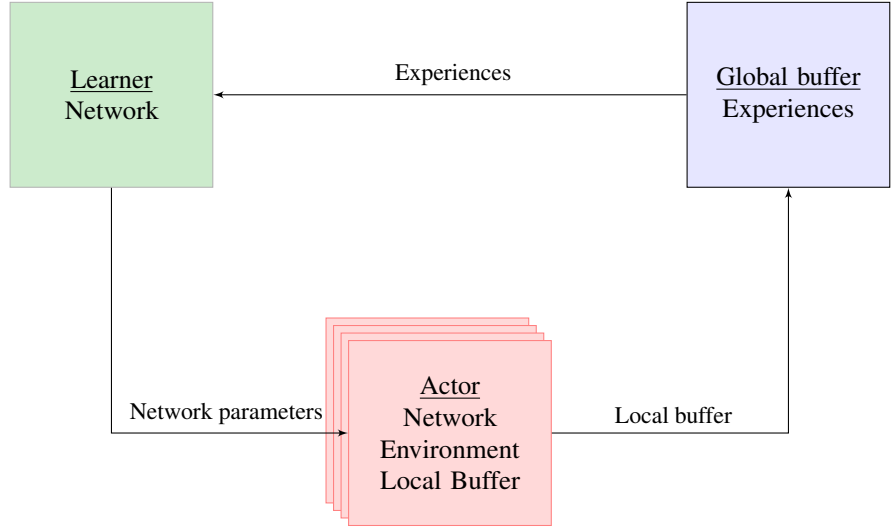
	
	The exploration requirements of the continuous state space, as defined previously introduce the necessity for collecting a large number of trajectories $M$. The total time required for gathering these trajectories heavily depends on the simulation time needed for one episode. In this particular setting developed, the simulation time can be quite long since, at each decision step, if the action selected is ``Trade", an optimisation model is constructed and solved.
	
	In this paper and in order to address this issue, we resort to an asynchronous architecture, similar to the one proposed in \cite{horgan2018distributed}, presented in Figure \ref{fig: Training}. The two processes, described in Section \ref{sec: Trajectories_Generation}, namely generation of trajectories and computation of the Q-functions, run concurrently with no high-level synchronization. 
	
	Multiple actors that run on different threads are used to generate trajectories. Each actor contains a copy of the environment, an individual $\epsilon$-greedy policy based on the latest version of the Q functions and a local buffer. The actors use their $\epsilon$-greedy policy to perform transitions in the environment. The transitions are stored in the local buffer. When the local buffer of each actor is filled, it is appended to the global buffer, the agent collects the latest Q-functions from the learner and continues the simulation. A single learner continuously updates the Q-functions using the simulated trajectories from a global buffer. 
	
	The benefits from asynchronous methods in Deep Reinforcement Learning (DRL) are elaborated in \cite{mnih2016asynchronous}. Each actor can use a different exploration policy (different initial $\epsilon$ value and decay) in order to enhance diversity in the collected samples which leads to a more stable learning process. Additionally, it is shown that the total computational time scales linearly with the number of threads considered. Another major advantage is that distributed techniques were shown to have a super-linear speedup for one-step methods that are not only related to computational gains. It is argued that, the positive effect of having multiple threads leads to a reduction of the bias in one-step methods \cite{mnih2016asynchronous}. In this way, these algorithms are shown to be much more data efficient than the original versions.

	\section{Case study}\label{sec: Results}
	The proposed methodology is applied for the case of a PHES unit. First, the parameters and the exogenous information used for the optimisation of the CID market participation of a PHES operator are described. Second, the benchmark strategy used for comparison purposes and the process that was carried out for validation are presented. Finally, performance results of the obtained policy are presented and discussed. 
	
	\subsection{Parameters specification}
	
	The proposed methodology is applied for a PHES unit participating in the German CID market with the following characteristics: 
	\begin{itemize}
		\item $\socmax{i} = 200$ MWh,
		\item $\socmin{i} = 0$ MWh,
		\item $SoC_{i}^{init} = SoC_{i}^{term} =\left(\socmax{i}-\socmin{i}\right) /2$,
		\item $\overline{C_{i}}=\overline{G_{i}}=200$ MW,
		\item $\underline{C_{i}}=\underline{G_{i}}=0$ MW, 
		\item $\eta = 100 \%.$
	\end{itemize}
	The discrete trading horizon has been selected to be ten hours, i.e. $T = \left\lbrace 17:00,...,03:00\right\rbrace$. The trading time interval is selected to be $\Delta t = 15$ min. Thus the trading process takes $K=40$ steps until termination. Moreover, all $96$ quarter-hourly products of the day, $X =\{Q_{1},..,Q_{96}\} $, are considered. Consequently, the delivery timeline is $\bar{T}=\left\lbrace 00:00,...,23:45 \right\rbrace$, with $\tau_{init}=00:00$ and $\tau_{term}=24:00$ and the delivery time interval is $\Delta \tau = 15$ min. Each product can be traded until 30 minutes before the physical delivery of electricity begins (e.g. $t_{close}(Q_{1})=23:30$ etc.). 
	
	The number of days selected for training was $|L^{train}| = 252$ days. For the construction of the training/test sets, the days were sampled uniformly without replacement from a pool of 362 days. The number of simulated episodes for each training day was selected to be $E=2000$ episodes for the artificial trajectories generation process, described in Section \ref{sec: Trajectories_Generation}. During the trajectories generation process the high-level actions (``Trade" or ``Idle") were chosen following an $\epsilon$-greedy policy. As described in Section \ref{sec: Async}, each of the actor threads is provided with a different exploration parameter $\epsilon$ that is initialised with a random uniform sample in the range $\left[ 0.1,0.5 \right]$. The parameter $\epsilon$ is then annealed exponentially until a zero value is reached.
	
	The pseudo-state $z_{i,t} = ( s'_{i,0},a'_{i,0},r_{i,0},...,a'_{i,t-1},r_{i,t-1},s'_{i,t} ) \in Z_{i}$ is composed of the entire history of observations and actions up to time-step $t$, as described in Section \ref{sec: SSR}. For the sake of memory requirements, as explained in Section \ref{sec: Network_architecture}, we assume that the last ten trading steps contain sufficient information about the past. Thus, the pseudo-state is transformed in sequences of fixed length $\bar{h}_{max}=10$, 
	
	\subsection{Exogenous variable}
	
	The exogenous variable $w^{exog}_{i,t}$ represents any relevant information available to agent $i$ about the system. In this case study, we assumed that the variable $w^{exog}_{i,t}$ contains:
	\begin{itemize}
		\item The 24 values of the Day-ahead price for the entire trading day
		\item The Imbalance price and the system Imbalance for the four quarters preceding each time-step $t$ 
		\item Time features: i) the hour of the day, ii) the month and iii) whether the traded day is a weekday or weekend
	\end{itemize}

	\subsection{Benchmark strategy}
	
	The strategy selected for comparison purposes is the ``rolling intrinsic" policy, denoted by $\pi^{RI}$ \cite{lohndorf2015optimal}. According to this policy, the agent selects at each trading time-step $t$ the action ``Trade", as described in Section \ref{sec: HLA}. This benchmark is selected since it represents the current state of the art used in the industry for the optimisation of PHES unit market participation.
	
	\subsection{Validation process} \label{sec: Validation_process}
	The performance of the policy obtained using the fitted Q algorithm, denoted by $\pi^{FQ}$, is evaluated on test set $L^{test}$ that contains historical data from 110 days. These days are not used during the training process. This process of backtesting a strategy on historical data is widely used because it can provide a measure of how successful a strategy would be if it had been executed in the past. However, there is no guarantee that this performance can be expected in the future. This validation process heavily relies on Assumption (\ref{asu: Rest_behaviour}) about the inability of the agent to influence the behaviour of the other players in the market. It can still provide an approximation on the results of the obtained policy before deploying it in real life. However, the only way to evaluate the exact viability of a strategy is to deploy it in real life. 
	
	We compare the performances of the policy obtained by the fitted Q iteration algorithm $\pi^{FQ}$ and the ``rolling intrinsic" policy $\pi^{RI}$. The comparison will always be based on the computation of the return of the policies on each day. For a given policy, the return over a day will be simply computed by running the policy on the day and summing up the rewards obtained.
	
	The learning algorithm that we have designed for learning a policy has two sources of variance, namely those related to the generation of the new trajectories and those related to the learning of the Q-functions from the set of trajectories. To evaluate the performance of our learning algorithm we will perform several runs and average the performances of the policies learned. In the following, when we report the performance of a fitted Q iteration policy over one day, we will actually report the average performances of ten learned policies over this day.

	We now describe the different indicators that will be used afterwards to assess the performances of our method. These indicators will be computed for both the training set and the test set, but are detailed hereafter when they are computed for the test set. It is straightforward to adapt the procedure for computing the indicators for the training set. 
	
	Let $V_{d}^{\pi^{FQ}}$ and $V_{d}^{\pi^{RI}}$ denote the total return of the fitted Q and the ``rolling intrinsic" policy for day $d$, respectively. We gather the obtained returns of each policy for each day $d\in L^{test}$. We sort the returns in ascending order, and we obtain an ordered set containing a number of $|L^{test}|$ values for each policy. We provide descriptive statistics about the distribution of the returns of each policy $V_{d}^{\pi^{FQ}}$ and $V_{d}^{\pi^{RI}}$ on the test set $L^{test}$. In particular, we report the mean, the minimum and maximum values achieved for the set considered. Moreover, we provide the values obtained for each of the quartiles (25\%, 50\% and 75\%) of the set. 
	
	Additionally, we compute the sum of returns over the entire set of days as follows:
	\begin{gather}
	V^{\pi^{FQ}} = \sum_{d\in L^{test}} V_{d}^{\pi^{FQ}} ,\label{eqn: CumReturnsOfFQpolicy}\\
	V^{\pi^{RI}} = \sum_{d\in L^{test}} V_{d}^{\pi^{RI}} .\label{eqn: CumReturnsOfRIpolicy}
	\end{gather}

	An alternative performance indicator considered is the discrepancy of the returns coming from the fitted Q policy with respect to the risk-averse rolling intrinsic policy. We define the profitability ratio $r_d$ for each day $d\in L^{test}$, that corresponds to the signed percentage difference between the two policies as follows:
	
	\begin{gather}
	r_d = \frac{V_{d}^{\pi^{FQ}}-V_{d}^{\pi^{RI}}}{V_{d}^{\pi^{RI}}} \cdot 100\% .
	\end{gather}
	
	In a similar fashion, we sort the profitability ratios obtained for each day in the test set and we provide descriptive statistics about its distribution across the set. The mean, minimum and maximum values of the profitability ratio as well as the values of each quartile are reported. Finally, we compute the profitability ratio for the sum of returns over the entire set between the two policies, as:
	
	\begin{gather}
	r_{sum} = \frac{V^{\pi^{FQ}}-V^{\pi^{RI}}}{V^{\pi^{RI}}} \cdot 100\% .
	\end{gather}
	\subsection{Results}
	The performance indicators described previously are computed for both the training and the test set. The results obtained are summarised in Tables \ref{table: Train_set_results} and \ref{table: Test_set_results}. Descriptive statistics about the distribution of the returns from both policies as well as the profitability ratio are presented for each dataset. 
	
	It can be observed that on average $\pi^{FQ}$ yields better returns than $\pi^{RI}$ both on the training and the test set. More specifically, on the training set, the obtained policy performs, on average $2.56\%$ better than the ``rolling intrinsic" policy. For the top $25\%$ of the training days the profitability ratio is higher than $3.69\%$ and in some cases it even exceeds $10\%$. Overall, the total profits coming from the fitted Q policy add up to \euro$2,167,624$, yielding a difference of \euro$57,423.3$ ($2.7\%$) more than the profits from the ``rolling intrinsic" for the set of 252 days considered.

	\begin{table}[h]
		\centering
		\caption{Descriptive statistics of the returns obtained on the days of the training set for policies $\pi^{FQ}$ and $\pi^{RI}$. The last column also provides the corresponding profitability ratios.}
		
		\label{table: Train_set_results}
		\begin{tabular}{cccc}
			& \textbf{$\pi^{FQ}$ returns (\euro)}& \textbf{$\pi^{RI}$ returns (\euro)}&\textbf{r (\%)} \\ \cmidrule[1pt]{1-4}
			$mean$ 				& $8670$		& $8440 $ & $2.56$ \\ 
			$min$ 				& $2586$ 		& $2524 $ & $-0.42$ \\
			$25\%$ 				& $5571$ 		& $5537 $ & $0.33$ \\
			$50\%$ 				& $7496$ 		& $7385 $ & $1.45$ \\
			$75\%$ 				& $11081$ 	& $10632$ & $3.69$ \\
			$max$ 				& $54967$ 	& $54285$ & $23.12$ \\
			$sum$				& $2167624 $ 	& $2110200$ & $ 2.7$
		\end{tabular}
	\end{table}

	The fitted Q policy yields on average a $1.5\%$ greater profit on the test set with respect to the returns of the ``rolling intrinsic" policy. It is important to highlight that for $50$\% of the test set, the profits from the fitted Q policy are higher than $~1$\% in comparison to the ``rolling intrinsic". The difference between the total profits resulting from the two policies over the set of 110 days considered amounts to \euro$15,681$ ($1.7\%$).
	
	\begin{table}[h]
		\centering
		\caption{Descriptive statistics of the returns obtained on the days of the test set for policies $\pi^{FQ}$ and $\pi^{RI}$. The last column also provides the corresponding profitability ratios.}
		
		\label{table: Test_set_results}
		\begin{tabular}{cccc}
			& \textbf{$\pi^{FQ}$ returns (\euro)}& \textbf{$\pi^{RI}$ returns (\euro)}&\textbf{r (\%)} \\ \cmidrule[1pt]{1-4}
			$mean$ 				& $8583$		& $8439$ & $1.50$ \\ 
			$min$ 				& $2391$ 	  & $2366$ & $-0.73$ \\
			$25\%$ 				& $5600 $ 	& $5527$ & $0.23$ \\
			$50\%$ 				& $7661 $ 	& $7622$ & $0.87$ \\
			$75\%$ 				& $10823$ 	& $10721$ & $2.22$ \\
			$max$ 				& $37902$ 	& $36490$ & $6.56$ \\
			$sum$				& $935552$ 	& $919871$ & $1.7$
		\end{tabular}
	\end{table}
	
\begin{figure}	
		\centering
		\hspace{-20pt}\includegraphics[width=0.5\textwidth]{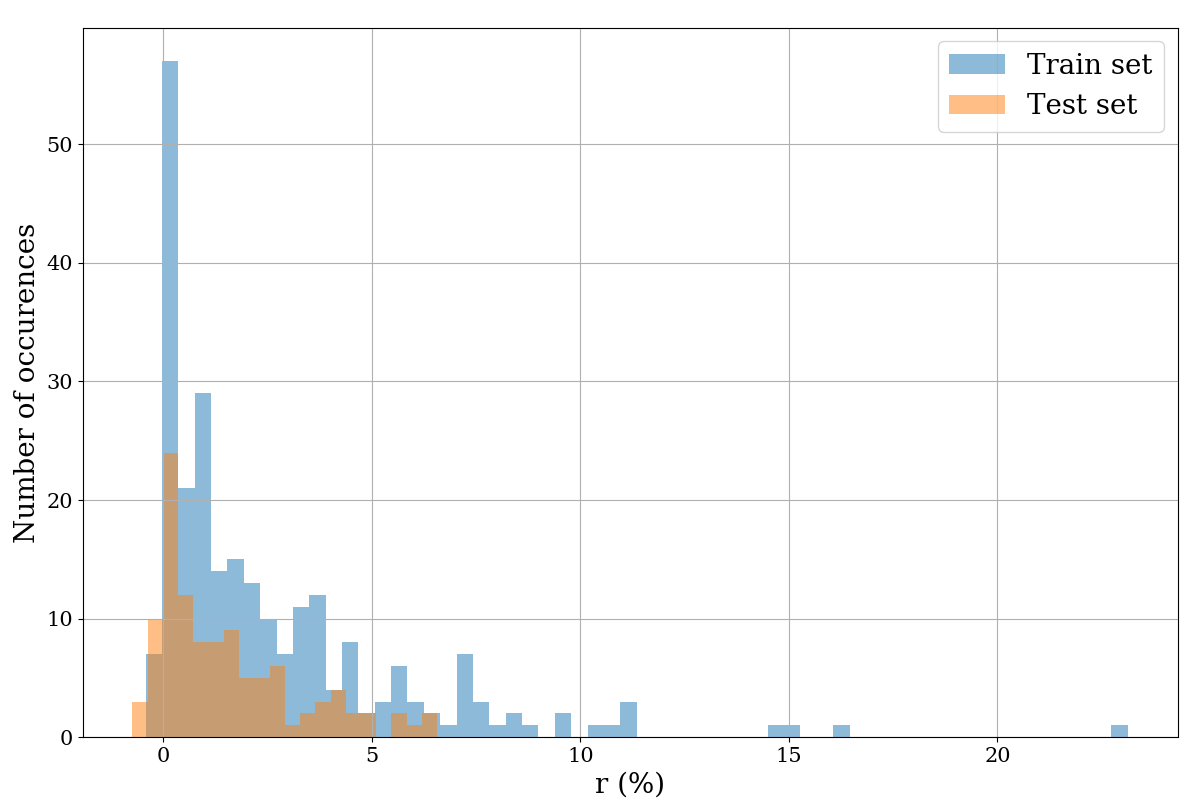}
		\caption{Profitability ratio.}
		\label{fig: Results}
	\end{figure}

	The distribution of training and test set samples according to the obtained profitability ratio is presented in Figure \ref{fig: Results}. It can be observed that most samples are spread in the interval between $0-5$\% and that the distribution has a positive skew. It is important to note that, for the vast majority of the samples, the profitability ratio is nonnegative. This result allows us to characterize the new policy as risk-free in the sense that in the worst case, the fitted Q policy performs as well as the ``rolling intrinsic" policy. From the standpoint of practical implementation this result allows us to construct a wrapper around the current industrial standard practices and expect an average improved performance of $2\%$ almost without any risk. However, as discussed earlier, the back-testing of a strategy in historical data may differ from the outcomes in real deployment for various reasons.

	\section{Discussion} \label{sec: Discussions}
	In this section, we provide some remarks related to the practical challenges encountered and the validity of the assumptions considered throughout this paper. 
	\subsection{Behaviour of the rest of the agents}
	
	In this paper, we assumed (Assumption \ref{asu: OtherAgents}) that the rest of the agents $-i$ post orders in the market based on their needs and some historical information of the state of the order book. In reality, the available information that the other agents possess is not accessible by agent $i$. This fact gives rise to issues related to the validity of the assumption that the process is Markovian. 
	
	We further assumed (\ref{asu: Rest_behaviour}) in Section \ref{sec: Trajectories_Generation} that the behaviour of agent $i$ does not influence the strategy of the other agents $-i$. Based on this assumption the training and the validation process were performed using historical data. However, the strategy of each of the market participants is highly dependent on the actions of the rest participants, especially in a market with limited liquidity such as the CID market. 
	
	These assumptions, although slightly unrealistic and optimistic, provide us with a meaningful testing protocol for a trading strategy. The actual profitability of a strategy can be obtained by deploying the strategy in real-time. However, it is important to show that the strategy is able to obtain substantial profits in back-testing first.

	\subsection{Partial observability of the process}
	
	In Section \ref{sec: RLProblem}, the decision-making problem studied in this paper was framed as an MDP after considering certain assumptions. Theoretically, this formulation is very convenient, but does not hold in practice. Especially considering Assumption \ref{asu: PseudoState}, where the reduced pseudo-state is assumed to contain all the relevant information required.
	
	Indeed, the trading agents do not have access to all the information required. For instance, a real agent does not know how many other agents are active in the market. They do not know the strategy of each agent either. There is also a lot of information gathered by $w^{exog}$ which is not available for the agent. Finally, the fact that the state space was reduced results in an inevitable loss of information.
	
	Therefore, it would be more accurate to consider a Partially Observable Markov Decision Process (POMDP) instead. In a POMDP, the real state is hidden and the agent only has access to observations. For an RL algorithm to properly work with a POMDP, the observations have to be representative of the real hidden states. 
	
	\subsection{Exploration}
	
	There are two main issues related to the state space exploration that result in the somewhat limited performance of the obtained policy. First, in the described setting, the way in which we generate the artificial trajectories is very important for the success of the method. The generated states must be ``representative" in the sense that the areas around these states are visited often under a near optimal policy \cite{bertsekas2005dynamic}. In particular, the frequency of appearance of these areas of states in the training process should be proportional to the probability of occurrence under the optimal policy. However, in practice, we are not in a position to know which areas are visited by the optimal policy. In that respect, the asynchronous distributed algorithm used in this paper was found to successfully address the issue of state exploration. 
	
	Second, the assumptions (Assumptions \ref{asu: StorageControl}, \ref{asu: Aggressor}, \ref{asu: Zero_Imbalances}) related to the operation of the storage device according to the ``default" strategy without any imbalances allowed, as well as the participation of the agent as an aggressor, are restrictive with respect to the set of all admissible policies. Additionally, the adoption of the reduced discrete action space described in Section \ref{sec: HLA} introduces further restrictions on the set of available actions. Although having a small and discrete space is convenient for the optimisation process, it leads to limited state exploration. For instance, the evolution of the state of charge of the storage device is always given as the output of the optimisation model based on the order book data. Thus, in this configuration, it is not possible to explore all areas of the state space (storage levels) but only a certain area driven by the historical order book data. However, evaluating the policy on a different dataset might lead to areas of the state space (e.g. storage level) that are never visited during training, leading to poor performance. Potential mitigations of this issue involve diverse data augmentation techniques and/or different representation of the action space.
	
	\section{Conclusions and future work}\label{sec: Conclusions}
	
	In this paper, a novel RL framework for the participation of a storage device operator in the CID market is proposed. The energy exchanges between market participants occur through a centralized order book. A series of assumptions related to the behaviour of the market agents and the operation of the storage device are considered. Based on these assumptions, the sequential decision-making problem is cast as an MDP. The high dimensionality of both the action and the state spaces increase the computational complexity of finding a policy. Thus, we motivate the use of discrete high-level actions that map into the original action space. We further propose a more compact state representation. The resulting decision process is solved using fitted Q iteration, a batch mode reinforcement learning algorithm. The results illustrate that the obtained policy is a low-risk policy that is able to outperform on average the state of the art for the industry benchmark strategy (``rolling intrinsic") by 1.5\% on the test set. The proposed method can serve as a wrapper around the current industrial practices that provides decision support to energy trading activities with low risk.
	
	The main limitations of the developed strategy originate from: i) the insufficient amount of relevant information contained in the state variable, either because the state reduction proposed leads to a loss of information or due to the unavailability of information and ii) the limited state space exploration as a result of the proposed high-level actions in combination with the use of historical data. To this end and as future work, a more detailed and accurate representation of the state should be devised. This can be accomplished by increasing the amount of information considered, such as RES forecasts, and by improving the order book representation. We propose the use of continuous high-level actions in an effort to gain state exploration without leading to very complex and high-dimensional action space. 
	
	\singlespacing
	
	\bibliography{journal}
	\bibliographystyle{icml2020}
	
	\appendix

	\section{Nomenclature}\label{sec: Appendix}
	
	\subsection*{\textbf{Acronyms}}
	\begin{tabularx}{\textwidth}{l X}
		ADP & Approximate Dynamic Programming.\\
		CID & Continuous Intraday.\\
		DRL& Deep Reinforcement Learning.\\
		FCFS & First Come First Served.\\
		MDP & Markov Decision Process. \\
		OB & Order Book. \\
		PHES & Pumped Hydro Energy Storage.\\
		RES & Renewable Energy Sources. \\
	\end{tabularx}
	
	\subsection*{\textbf{Sets and indexes}}
	\begin{supertabular}{l p{0.8\columnwidth}}
		Name & Description \\
		\hline
		$i$ & Index of an agent. \\
		$-i$ & Index of all the agents except agent $i$. \\
		$j$ & Index of an order. \\
		$m$ & Index of a sample of quadruples. \\
		$d$ & Index of a day in a set. \\
		$t$ & Trading time-step. \\
		$\tau$ & Discrete time-step of delivery.\\
		
		$A$ & Joint action space for all the agents.\\
		$A_{i}$ & Action space of agent $i$.\\
		$A_{-i}$ & Action space of the rest of the agents $-i$.\\
		$A_{i}^{red}$ & Reduced action space of agent $i$.\\
		$A'_{i}$& Set of high-level actions for agent $i$.\\
		$\bar{A}_{i}$ & Set of all factors for the partial/full acceptance of orders by agent $i$.\\
		$ E $ & Set of conditions that can apply to an order.\\
		$F$ & Set of all sampled trajectories.\\
		$F'$ & Set of sampled one-step transitions.\\
		$F'_t$ & Set of sampled one-step transitions for time $t$.\\
		$ H_{i} $ & Set of all histories for agent $i$.\\
		$I$ & Set of agents. \\
		$L^{train}$ & Set of trading days used to train the agent.\\
		$L^{test}$ & Set of trading days used to evaluate the agent.\\
		$ N_t $ & Set of all available order unique indexes at time $t$.\\
		$ N'_t $ & Set of all the unique indexes of new orders posted at time $t$.\\
		$ N_{\tau} $ & Set of all the unique indexes of orders for delivery at $\tau$.\\
		$ O_t $ & Set of all available orders in the order book at time $t$.\\
		$ S^{OB}$ & Set of all available orders in the order book. \\
		$ S'^{OB}$ & Low dimensional set of all available orders in the order book. \\
		$ S_{i}$ & State space of agent $i$. \\
		$ T $ & Trading horizon, i.e. time interval between first possible trade and last possible trade.\\
		$ T(x) $ & Discretization of the trading timeline for product $x$.\\
		$ \bar{T} $ & Discretization of the delivery timeline.\\
		$ \bar{T}(t) $ & Discretization of the delivery timeline at trading step $t$.\\
		$ T^{Imb} $ & Discretization of the imbalance settlement timeline.\\
		$ X $ & Set of all available products.\\
		$ X_t $ & Set of all available products at time $t$.\\
		$ Z_{i}$& Set of pseudo-states for agent $i$.\\
		$ \Pi $ & Set of all admissible policies.\\
	\end{supertabular}
	
	\subsection*{\textbf{Parameters}}
	\begin{supertabular}{l p{0.8\columnwidth}}
		Name & Description \\
		\hline
		
		$\overline{C_{i}}$ & Maximum consumption level for the asset of agent $i$.\\ 
		$\underline{C_{i}}$ & Minimum consumption level for the asset of agent $i$.\\ 
		$E$ & Number of episodes.\\
		$e$ & Conditions applying on an order other than volume and price.\\
		$ep$ & Number of simulations between two successive Q function updates.\\
		$decay$ & Parameter for the annealing of $\epsilon$.\\
		$\overline{G_{i}}$ & Maximum production level for the asset of agent $i$.\\ 
		$\underline{G_{i}}$ & Minimum production level for the asset of agent $i$.\\
		$\bar{h}$ & Sequence length of past information.\\
		$\bar{h}_{max}$& Maximum sequence length of past information.\\
		$I(\tau)$ & Imbalance price for delivery period $\timespan{x}$.\\
		$K$ & Number of steps in the trading period.\\
		$M$ & Number of samples of quadruples. \\
		$n$ & Number of agents. \\
		$o_t$ & Market order. \\
		$p$ & Price of an order.\\
		$p_{\max}$ & Maximum price of an order. \\
		$p_{\min}$ & Minimum price of an order. \\
		$\socmax{i}$ & Maximum state of charge of storage device.\\ 
		$\socmin{i}$ & Minimum state of charge of storage device.\\ 
		$SoC_i^{init}$ & State of charge of storage device at the beginning of the delivery timeline.\\ 
		$SoC_i^{term}$ &  State of charge of storage device at the end of the delivery timeline.\\
		$ t_{close}(x)$ & End of trading period for product $x$.\\
		$ t_{delivery}(x)$ & Start of delivery of product $x$.\\
		$ t_{open}(x)$ & Start of trading period for product $x$.\\
		$ t_{settle}(x)$ & Time of settlement for product $x$.\\
		$v$ & Volume of an order.\\
		$x$ & Market product. \\ 
		$y$ & Side of an order (``Sell" or ``Buy"). \\
		$y^{m}_t$ & Target computed for sample $m$ at time $t$.\\
		$\timespan{x}$ & Time interval covered by product $x$ (delivery).\\
		$\Delta t $ & Time interval between trading time-steps.  \\
		$\Delta \tau $ & Time interval between  delivery time-steps.  \\
		$\epsilon$ & Parameter for the $\epsilon$-greedy policy. \\
		$\eta$ & Charging/discharging efficiency of storage device.\\
		$\theta_t $ & Parameters vector of function approximation at time $t$.  \\
		$\lambda(x)$ & Duration of time-interval $\timespan{x}$.\\
		$\zeta$ & A single trajectory. \\
		$\zeta_{m}$ & A single indexed trajectory. \\
		$\tau_{init}$& Initial time-step of the delivery timeline. \\
		$\tau_{term}$& Terminal time-step of the delivery timeline. \\
	\end{supertabular}
	
	\subsection*{\textbf{Variables}}
	\begin{supertabular}{l p{0.8\columnwidth}}
		Name & Description \\
		\hline
		$a_{t}$ & Joint action from all the agents at time $t$.\\
		$a_{i,t}$ & Action of posting orders by agent $i$ at time $t$.\\
		$a_{-i,t}$ & Action of posting orders by the rest of the agents $-i$ at time $t$.\\
		$a'_{i,t}$ & High-level action by agent $i$ at time $t$.\\
		$a_{i,t}^{j}$ & Acceptance (partial/full) factor for order $j$ by agent $i$ at time $t$.\\
		$\bar{a}_{i,t}$ & Factors for the partial/full acceptance of all orders by agent $i$ at time $t$.\\
		$C_{i,t}(\tau)$ & Consumption level at delivery time-step $\tau$ computed at time $t$.\\
		$c_{i,t}(t')$ & Consumption level during the delivery interval.\\
		$e_{i,t}$& Random disturbance for agent $i$ at time $t$.\\
		$G_{i,t}(\tau)$ & Generation level at delivery time-step $\tau$ computed at $t$.\\ 
		$g_{i,t}(t')$ & Generation level during the delivery interval.\\
		$h_{i,t}$ & History vector of agent $i$ at time $t$.\\
		$k_{i,t}(\tau)$& Binary variable that enforces either charging or discharging of the storage device.\\
		$P_{i,t}^{mar}(x)$ & Net contracted power of agent $i$ for product $x$ (delivery time-step $\tau$) at time $t$.\\
		$P^{res}_{i,t}(\tau)$ & Residual production of agent $i$ delivery time-step $\tau$ (for product $x$) at time $t$.  \\
		$P^{res}_{i}(\tau)$ & Final residual production of agent $i$ for product\\
		$r_{i,t}$ & Instantaneous reward of agent $i$ at time $t$.\\
		$r_{d}$ & Profitability ratio at day $d$.\\
		$r_{sum}$ & Profitability ratio for the sum of returns over set.\\
		$s_{i,t}$ & State of agent $i$ at time $t$. \\
		$\soc{i}{t}(\tau)$ & State of charge of device at delivery time-step $\tau$ computed at $t$.\\
		$s^{OB}_t$ & State of the order book at time $t$.\\
		$s'^{OB}_t$ & Low dimensional state of the order book at time $t$.\\
		$s^{private}_{i,t}$ & Private information of agent $i$ at time $t$.\\
		$\bar{s}_t$& Triplet of fixed size, part of pseudo-state $z_{i,t}$ that serves as an input at LSTM at time $t$.\\
		$u_{i,t}$ & Aggregate (trading and asset) control action of the asset trading agent $i$ at time $t$.\\
		$v^{con}_{i,t}(x)$ & Volume of product $x$ contracted by agent $i$ at time $t$.\\
		$w^{exog}_{i,t}$ & Exogenous information of agent $i$ at time $t$.\\
		$z_{i,t}$ & Pseudo-state for agent $i$ at time $t$.\\
		$\Delta_{i,t}(\tau)$ & Imbalance for delivery time $\tau$ for agent $i$ computed at time $t$.\\
		$\Delta_{i}(\tau)$ & Final imbalance for delivery time $\tau$ for agent $i$.\\
		$\Delta G_{i,t}$ & Change in the production level for the asset of agent $i$ at time $t$.\\ 
		$\Delta C_{i,t}$ : & Change in the consumption level for the asset of agent $i$ at time $t$.\\

	\end{supertabular}
	
	\subsection*{\textbf{Functions}}
	\begin{supertabular}{l p{0.8\columnwidth}}
		Name & Description \\
		\hline	
		$clear(\cdot)$ & Market clearing function. \\
		$b(\cdot)$ & Univariate stochastic model for exogenous information. \\
		$e(\cdot)$ & Encoder that maps from the original state space $H_{i}$ to pseudo-state space $Z_{i}$. \\
		$f(\cdot)$ & Order book transition function. \\
		$G^{\zeta}(\cdot)$ & Revenue collected over a trajectory. \\
		$g(\cdot)$ & System dynamics of the MDP. \\
		$k(\cdot)$ & System dynamics of asset trading process. \\
		$l(\cdot)$ & Reduced action space construction function. \\
		$P_{a_{-i,t}(\cdot)}$ & Probability distribution function for the actions of the rest of the agents $-i$. \\
		$P_{e_t}(\cdot)$ & Random disturbance probability distribution function. \\
		$P(\cdot)$ & Transition probabilities of the MDP. \\
		$P_{FQ}(\cdot)$ & The stochastic process (algorithm) of fitted Q iteration. \\
		$P_{\theta_{t,0}}(\cdot)$ & Distribution of the initial parameters $\theta_{t,0}$. \\
		$p(\cdot)$ & Mapping from high-level actions $A'_{i}$ to the reduced action space $A_{i}^{red}$. \\
		$Q_{t}(\cdot, \cdot)$ & State-action value function at time $t$. \\
		$\hat{Q}(\cdot, \cdot)$ & Sequence of Q-function approximations. \\
		$R(\cdot)$ & Reward function. \\
		$u(\cdot)$ & Signing convention for the volume wrt. the side (`Buy" or `Sell") of each order. \\
		$V^{\pi_{i}}(\cdot)$ & Total expected reward function for policy $\pi_i$.\\
		$V^{\pi^{FQ}}_{d}(\cdot)$ & Return of the fitted Q policy $\pi^{FQ}_i$ for day $d$.\\
		$V^{\pi^{RI}}_{d}(\cdot)$ & Return of the  ``rolling" intrinsic policy $\pi^{RI}_i$ for day $d$.\\
		$\mu_{t} (\cdot)$ & Policy function at time $t$.\\
		$\pi_i(\cdot)$ & Policy followed by agent $i$.\\
		$\rho(\cdot)$ & Trading revenue function. \\
		
	\end{supertabular}

\end{document}